\documentclass[review]{elsarticle}

\usepackage{ulem}
\usepackage[table,xcdraw]{xcolor}
\usepackage{soul}
\usepackage{colortbl}
\usepackage{makecell}
\usepackage{url}
\usepackage{multirow}

\usepackage{tikz}
\usetikzlibrary{arrows,positioning}
\usetikzlibrary{plotmarks}
\usetikzlibrary{calc}
\usetikzlibrary{shapes.geometric}
\usetikzlibrary{decorations.pathreplacing,calligraphy}

\newcommand{\SpearmanCoefficient}{\rho}

\journal{Journal}

\biboptions{authoryear}
\begin{document}

\begin{frontmatter}

% \title{End-to-end deep learning pipeline for  bidimensional and volumetric tumor burden measurement from MRI in pre and postoperative GBM patients}

\title{Deep learning automates bidimensional and volumetric tumor burden measurement from MRI in pre- and post-operative glioblastoma patients}

%% Group authors per affiliation:
\author{Jakub Nalepa$^{1,2}$}\ead{Jakub.Nalepa@polsl.pl}\corref{mycorrespondingauthor}
\author{Krzysztof Kotowski$^{1}$}
\author{Bartosz Machura$^{1}$}
\author{Szymon Adamski$^{1}$}
\author{Oskar Bozek$^{3}$}
\author{Bartosz Eksner$^{4}$}
\author{Bartosz Kokoszka$^{4}$}
\author{Tomasz Pekala$^{5}$}
\author{Mateusz Radom$^{6}$}
\author{Marek Strzelczak$^{6}$}
\author{Lukasz Zarudzki$^{6}$}
\author{Agata Krason$^{7}$}
\author{Filippo Arcadu$^{8}$}
\author{Jean Tessier$^{7}$}
\address{
$^1$Future Processing Healthcare, Gliwice, Poland\\
$^2$Department of Algorithmics and Software, Silesian University of Technology, Gliwice, Poland\\
$^3$Department of Radiodiagnostics and Invasive Radiology, School of Medicine in Katowice, Medical University of Silesia in Katowice, Katowice, Poland\\
$^4$Department of Radiology and Nuclear Medicine, ZSM Chorzów, Chorzów, Poland\\
$^5$Department of Radiodiagnostics, Interventional Radiology and Nuclear Medicine, University Clinical Centre, Katowice, Poland\\
$^6$Department of Radiology and Diagnostic Imaging, Maria Skłodowska‑Curie National Research Institute of Oncology, Gliwice Branch, Gliwice, Poland\\
$^7$Roche Pharmaceutical Research \& Early Development, Early Clinical Development Oncology, Roche Innovation Center Basel, Basel, Switzerland\\
$^8$Roche Pharmaceutical Research \& Early Development, Early Clinical Development Informatics, Roche Innovation Center Basel, Basel, Switzerland}
\cortext[mycorrespondingauthor]{Corresponding author}

%\fntext[myfootnote]{Since 1880.}

%% or include affiliations in footnotes:
% \author[mymainaddress,mysecondaryaddress]{Elsevier Inc}
% \ead[url]{www.elsevier.com}

% \author[mysecondaryaddress]{Global Customer Service\corref{mycorrespondingauthor}}
%\cortext[joint]{Joint first author}
% \address[mymainaddress]{1600 John F Kennedy Boulevard, Philadelphia}
% \address[mysecondaryaddress]{360 Park Avenue South, New York}

\begin{abstract}

Tumor burden assessment by magnetic resonance imaging (MRI) is central to the evaluation of treatment response for glioblastoma. This assessment is complex to perform and associated with high variability due to the high heterogeneity and complexity of the disease. In this work, we tackle this issue and propose a deep learning pipeline for the fully automated end-to-end analysis of glioblastoma patients. Our approach simultaneously identifies tumor sub-regions, including the enhancing tumor, peritumoral edema and surgical cavity in the first step, and then calculates the volumetric and bidimensional measurements that follow the current Response Assessment in Neuro-Oncology (RANO) criteria. Also, we introduce a rigorous manual annotation process which was followed to delineate the tumor sub-regions by the human experts, and to capture their segmentation confidences that are later used while training the deep learning models. The results of our extensive experimental study performed over 760 pre-operative and 504 post-operative adult patients with glioma obtained from the public database (acquired at 19 sites in years 2021--2020) and from a clinical treatment trial (47 and 69 sites for pre-/post-operative patients, 2009--2011) and backed up with thorough quantitative, qualitative and statistical analysis revealed that our pipeline performs accurate segmentation of pre- and post-operative MRIs in a fraction of the manual delineation time (up to 20 times faster than humans). The bidimensional and volumetric measurements were in strong agreement with expert radiologists, and we showed that RANO measurements are not always sufficient to quantify tumor burden.

\end{abstract}

\end{frontmatter}

%\linenumbers

\section{Introduction}\label{sec:introduction}

Glioblastoma (GBM) is the most common of malignant primary brain tumors in adults and despite decades of research, still remains one of the most feared of all cancer types due to its poor prognosis. Accurately evaluating response to therapies in GBM also presents considerable challenges. It is based on the use of the Response Assessment in Neuro-Oncology (RANO) criteria and the measurement of two perpendicular diameters of the contrast-enhancing tumor (ET) area~\citep{Ellingson2017}, as well as a qualitative evaluation of abnormalities on T2-weighted and FLAIR MRI sequences, which correspond to regions of edema (ED) with or without tumor cell infiltration. Radiological evaluation is notoriously complex as the tumors are often very heterogeneous in appearance, with an irregular shape associated with the infiltrative nature of the disease. The effect is further compounded in the post-surgical setting, by the presence of the surgical cavity and brain distortion. Inevitably, this leads to high intra- and inter-reader variability which, in turn, limits our ability to detect a therapeutic benefit in clinical trials and capture early patient response or progression in clinical practice. In this work, we tackle this issue and propose an end-to-end deep learning pipeline for the segmentation of
tumor sub-regions from MRI for patients with glioma and the subsequent automated measurements of their volumetric characteristics and bidimensional diameters---our most important contributions are summarized in Section~\ref{sec:contribution}, whereas Section~\ref{sec:related_work} presents the related literature.

\subsection{Related literature}\label{sec:related_work}

Segmentation of glioblastoma (former name: glioblastoma multiforme) from multi-sequence MRI scans has been extensively researched in the literature~\citep{bakas2019identifying, NALEPA2020101769}, and it was popularized by the Brain Tumor Segmentation (BraTS) challenge that is focusing on developing automated algorithms for accurate brain tumor multi-class segmentation~\citep{Menze2015, baid2021rsnaasnrmiccai}. The existing algorithms targeting the brain tumor detection and segmentation are commonly split into three main categories~\citep{liu_survey_2014, wadhwa_review_2019}---(\textit{i})~atlas-, (\textit{ii})~image analysis-, and (\textit{iii})~machine learning-based techniques. Such approaches may be also assembled into hybrid techniques which benefit from the advantages of the combined algorithms~\citep{ZHAO201898,Habib2021,Kadry2021}.

The atlas-based models~\citep{10.3389/fnins.2020.00585}, that span across single- and multi-atlas label propagation techniques, segment the input MRI scans by extrapolating them to the previously curated reference images (referred to as \textit{atlases}) that represent the natural anatomical variability of the brain tissue~\citep{park_derivation_2014}. This is often a two-step process involving the global image registration that initially aligns the scans, and fine-tuning that captures local adaptations to specific brain anatomies (there exist methods that combine non-rigid registration with various tumor growth models~\citep{5627302}). Therefore, the overall quality of tumor segmentation is directly influenced by the representativeness of the utilized atlases (which are not necessarily transferable across scans of different characteristics~\citep{ALJABAR2009726}) and the performance of the co-registration process, thus developing well-performing deformable image registration techniques is of high practical importance in this context~\citep{MOHAMED2006752}. The latter approaches can also lead to non-rigid transformations which are not necessarily anatomically plausible, especially in the case of deformed brains~\citep{NALEPA2020101769}. Such methods are, however, easy to parallelize and interpret, but require building comprehensive atlases in the cumbersome, costly, and user-dependent process. This procedure could be accelerated through employing semi-automated techniques~\citep{SAGBERG2019101658}, especially in specific scenarios, e.g.,~for post-operative patients with affected brain anatomies.

Classical image analysis algorithms are usually classified into (\textit{i})~the approaches that label the voxels based on their intensity through employing various (single- and multi-level~\citep{Al-Rahlawee2021}) thresholding approaches~\citep{ILHAN2017580} or (\textit{ii})~the algorithms that analyze the characteristics of the voxel's neighborhood in a region-based manner. The former techniques offer real-time operation and are trivial to implement, but the appropriate threshold values should be either fine-tuned beforehand, or dynamically selected based on the input MRI characteristics. Additionally, since the brain tumors may manifest significant intensity variations, thresholding-based approaches can fail for heterogeneous lesions, and are susceptible to image noise, hence commonly require additional de-noising~\citep{Sharif2018}. On the other hand, exploiting the neighborhood voxel's information can robustify such techniques through extracting hand-crafted features. In region growing algorithms, the input scan is split into coherent regions, according to specific similarity metrics that should be selected \textit{a priori}~\citep{SrinivasaReddy2021}. In the active contour approaches, the initially determined contours are evolved toward the exact tumor boundaries~\citep{ESSADIKE2018103}. Such techniques track the tumor boundaries through matching a deformable model to the object (lesion), according to the energy functional that effectively controls the rigidity and elasticity of the curve~\citep{https://doi.org/10.1002/ima.22205,10.1371/journal.pone.0183943}. Although being thoroughly researched in the computer vision and medical image analysis fields, active contour techniques are still heavily parameterized and do not perform well in the case of sharp corners, concavities, and smooth boundaries~\citep{NALEPA2020101769}. They are, however, commonly used in semi-supervised segmentation pipelines, in which practitioners can provide valid initial contouring and parameterizations of the deformable model~\citep{10.1117/12.2042915}.

Machine learning approaches for brain tumor segmentation are split into the (\textit{i})~conventional techniques which require manual feature engineering, including feature extraction often followed by feature selection which aims at determining a subset of the most discriminative image features~\citep{9091724,VarunaShree2018,Abbas_2021}, and (\textit{ii})~deep learning algorithms that learn features from the data automatically during the training process, hence benefit from automated representation learning~\citep{NASER2020103758}. In unsupervised segmentation algorithms~\citep{WU2021242}, we do not exploit manually-delineated training sets, but utilize the data characteristics---captured in the input or specific feature spaces~\citep{10.3389/fnins.2021.634926}---to partition the image data into consistent clusters of voxels~\citep{Chen2018,WU2021242}. Unsupervised segmentation can be conveniently exploited as a pre-processing step, e.g.,~while generating ground-truth examples that could be later used for training supervised learners, as the pre-segmented MRI scans are much faster to analyze, interpret, and ultimately annotate by human readers. This partial automation which is offered by unsupervised learning has been recently used in the unsupervised quality control of segmentations, where the quality estimates are produced by comparing each segmentation with the output of a probabilistic segmentation model that relies on certain intensity and smoothness assumptions. Here, unsupervised approach helps determine atypical segmentation maps, and even predict the performance of segmentation algorithms~\citep{AUDELAN2021101895}. On the other hand, supervised learners utilize the ground-truth delineations in the training process. Such algorithms span across numerous well-established and widely utilized classification engines, and include, among others, support vector machines~\citep{Mishra2021}, random forests~\citep{10.1007/978-3-319-55524-9_9}, $k$-nearest neighbor ($k$-NN) classifiers~\citep{Kirtania2020}, extreme learning machines~\citep{Sasank2021}, ensemble learners~\citep{Barzegar2020}, and others~\citep{https://doi.org/10.1002/mp.13649}. Although some of the aforementioned models are trivial to interpret, e.g.,~$k$-NNs, it is not the case for non-linear algorithms which additionally require fine-tuning their pivotal hyper-parameters, and they can be sensitive to noise in the training sample. Therefore, selecting appropriate training sets can play a critical role in elaborating well-generalizing machine learning models~\citep{Nalepa2019}.

Currently, deep learning has been blooming in the field of automatic analysis of medical image data, and has established the state of the art in numerous segmentation tasks through delivering the winning solutions in the recent biomedical segmentation competitions~\citep{Isensee2021nnunet,SALEEM2021104410}. Automated representation learning that underpins the deep learning algorithms allows us to reveal features that may be impossible to capture by humans---there have been a multitude of deep architectures proposed so far for automating the process of brain tumor contouring. Such approaches include a variety of convolutional neural networks~\citep{BENNACEUR2020101692,10.3389/fncom.2019.00056}, generative adversarial models~\citep{YUAN2020101731}, residual architectures~\citep{Saha2021}, context-aware models~\citep{Pei2020}, inception-based networks~\citep{10.3389/fncom.2019.00044}, and many more~\citep{TAJBAKHSH2020101693}, but the recent BraTS editions clearly indicate that the U-Net-based~\citep{ronneberger_u-net_2015} architectures outperform other approaches for this task. The variations of this encoder-decoder model encompass multi-level cascaded techniques which sequentially detect whole-tumor areas, and then segment specific types of the brain tumor tissue~\citep{DBLP:conf/brainles-ws/KotowskiAMMZN20}, lightweight U-Nets~\citep{DBLP:conf/brainles-ws/TarasiewiczKN20}, the U-Nets with various loss functions, also capturing the boundary tumor's characteristics~\citep{DBLP:conf/brainles-ws/LorenzoMN19}, hardware-optimized models~\citep{Xiong2021}, hybrid algorithms that combine U-Nets with densely-connected and residual architectures~\citep{Zeineldin2020}, ensembled U-Nets~\citep{10.3389/fradi.2021.704888}, and many more~\citep{9103502,10.1145/3450519,ZHANG2021107562,ZHANG2021195}. Building a new deep learning algorithm for a given task requires defining not only the network's architecture, but also its most hyper-parameters~\citep{DBLP:conf/gecco/LorenzoNKRP17}, such as its depth, number and size of kernels, data augmentation routines executed to synthesize training data if necessary~\citep{8803423,Shorten2019,10.3389/fncom.2019.00083}, pre- and post-processing operations, or training strategy. \cite{Isensee2021nnunet}~approached this issue and introduced the nnU-Net framework that allows us to automatically optimize U-Net-based models for medical image segmentation. The recent BraTS editions showed that nnU-Nets became the model of choice for accurate brain tumor segmentation---they not only won BraTS 2020 with a significant margin~\citep{10.1007/978-3-030-72087-2_11} (and other biomedical challenges~\citep{Isensee2021nnunet}), but are commonly deployed in emerging algorithms. In this work, we follow this research pathway and build upon the nnU-Net framework and extend it through the exploitation of the radiologists' confidence concerning their manual annotations during the training process. Of note, we have thoroughly investigated 34 deep learning architectures to fully understand the impact of the most important architectural choices, together with the selection of training data on the abilities of the U-Net algorithms.

Overall, there exist a plethora of brain tumor delineation algorithms, but segmentation is virtually never the final step in the processing chain---the segmented areas are further analyzed to extract quantifiable tumor's characteristics (thus, inaccurate segmentation will directly affect the quality of all other steps). However, manual analysis if often affected by the subjectivity of the rater leading to high intra- and inter-rater variability~\citep{10.3389/fpsyg.2017.01628,VISSER2019101727}, which adversely impact our capabilities of objectively monitoring the patient's response to the treatment in longitudinal studies, clinical trials, and multi-center investigations. Therefore, automating the entire analysis chain is of utmost practical importance, as it could allow us to extract patient-specific information in a fully reproducible and unbiased way~\citep{NALEPA2020101769}. Virtually all computer-assisted methods that target GBM patients exploit semi-automated techniques~\citep{Egger2013}, in which---although significantly accelerated---the interaction with the user is still an important factor affecting quality and reproducibility of the calculated measures, e.g.,~volumetric tumor's characteristics~\citep{Chow498,Sezer2020}. \cite{Meier2017Autom} showed that automatic estimation of extent of resection and residual tumor volume of patients with GBM are comparable to the estimates of human experts, but commonly lead to overestimated volumes. The end-to-end fully-automated pipeline that we propose in this paper builds upon the pioneering work done by \cite{Chang2019GBM} in post-operative GBM, recently used in pediatric brain tumors~\citep{Peng2021Children}, with significant improvements that allow us to address the limitations of the aforementioned work, and related to the segmentation and RANO bidimensional calculation algorithms, clinical data preparation, assessment and utilization, and finally verification over a large and heterogeneous multi-center clinical data (the detailed comparison with \cite{Chang2019GBM} is given in the Supplementary Material). To our knowledge, the work presented here is the first reported deep learning algorithm capable of identifying and segmenting multiple tumor areas for both pre and post-surgery patients with glioma. Additionally, it objectively assesses the tumorous regions by extracting their quantifiable volumetric and bidimensional characteristics.

\subsection{Contribution}\label{sec:contribution}

In this paper, we report on our effort to ease the radiological interpretation of the disease with the development of an end-to-end deep learning pipeline for the segmentation of brain tumors and subsequent automated measurements of their volumes and bidimensional diameters. We thoroughly investigate the performance of our technique and its ability to discriminate and quantify tumor sub-regions in pre-surgery as well as in post-surgery settings by comparing the automated assessment to the results obtained by expert radiologists. Overall, our contributions are multi-fold and can be summarized by the following bullet points:
\begin{itemize}
    \item We propose an end-to-end pipeline for the simultaneous segmentation of the brain tumor sub-regions (enhancing tumor, peritumoral edema and surgical cavity) from multi-modal MRI (Section~\ref{sec:algorithm})\footnote{It is worth mentioning that we based on the segmentation algorithm introduced in this work in the newest (and largest so far, encompassing the MRI data of more than 2000 patients from 37 institutions) edition of the BraTS Challenge which allowed us to take the 6$^{\rm th}$ place out of 1600 participants (for details, see \url{https://www.rsna.org/news/2021/november/2021-AI-Challenge-Winners}; last access: December 8, 2021.}. The proposed approach benefits from the recent advances in the deep learning field and builds upon and extends the nnU-Net engine which established the state of the art in the medical image segmentation for a variety of organs and modalities~\citep{Isensee2021nnunet,SALEEM2021104410}. Once the sub-regions are delineated, we automatically measure the bidimensional and volumetric characteristics of the tumor, according to the current RANO criteria. Additionally, we introduce a new way of calculating RANO which is less sensitive to small alterations in the contour of the lesions, hence more robust against varying-quality automated segmentations and jagged contouring.
    \item We introduce a rigorous manual annotation procedure (Section~\ref{sec:manual_procedure}) and follow it in this work to ensure the high quality of manual delineations, and to capture additional information related to the segmentation process, including the time of manual analysis alongside the readers' confidences concerning each tumor sub-region. The confidences are utilized in the proposed confidence-aware nnU-Nets during their training process.
    \item We perform an extensive multi-fold experimental validation of the proposed pipeline in order to fully understand and quantify its capabilities in a range of clinical settings (Section~\ref{sec:results}). We utilized 760 pre-operative and 504 post-operative MRIs captured in dozens of institutions, therefore they manifest high level of data heterogeneity (Section~\ref{sec:data}). The experimental results indicate that the algorithm is able to analyze MRI scans for both pre-operative and post-operative patients producing results that are often more consistent, accurate and reliable than human experts in a fraction of time required by expert radiologists (up to 20$\times$ faster). To our knowledge, our technique is the first in the literature that effectively targets both pre- and post-operative MRIs using a single deep learning algorithm, and has been validated over a large and heterogeneous set of MRI data.
    % \item We present the detailed architecture of our model which makes it reproducible over other MRI data.
\end{itemize}

\section{Materials and methods}\label{sec:materials}

\subsection{Patient cohorts}\label{sec:data}

We collected five cohorts of adult patients with glioma consisting of MRI visits either in pre-surgery (4 cohorts) or post-surgery setting (1 cohort) (Table~\ref{tab:cohorts}). The Brain Tumor Segmentation (BraTS 2020) pre-operative cohorts consisted of 660 patients with pre-operative MRI scans of GBM/high-grade gliomas (HGGs) and low-grade gliomas (LGGs) with pathologically confirmed diagnosis, captured in 19 institutions using different equipment and imaging protocols (years 2012--2020)~\citep{Bakas2017,BRATS1,BRATS2, Menze2015}. This dataset was divided into a training set BraTS 2020~(Tr) of 369 patients, a validation set BraTS 2020~(V) of 125 patients and a test set BraTS 2020~(Te) of 166 patients---the brain tumor labels are publicly available only for BraTS 2020~(Tr). To the best of our knowledge, BraTS is the largest and the most comprehensive public domain GBM MRI dataset (in terms of types of utilized scanners, imaging protocols and types of imaged lesions) currently available for validating emerging tumor segmentation algorithms, such as within the framework of the Brain Tumor Segmentation Challenge organized yearly~\citep{bakas2019identifying,baid2021rsnaasnrmiccai}. The MRIs include the native T1 (originally acquired in sagittal or axial orientation), post-contrast T1 (Gadolinium, 3D axial), T2-weighted (2D axial) and FLAIR (2D axial, coronal or sagittal) sequences (all with variable slice thickness) co-registered to a common anatomical template~\citep{Torsten2010}, skull-stripped and resampled to 1~mm$^3$. The data is publicly available through the Image Processing Portal of the Center for Biomedical Image Computing and Analytics (CBICA) at the University of Pennsylvania, USA (\url{https://ipp.cbica.upenn.edu/}; last access: December 7, 2021).

\begin{table}[ht!]
\scriptsize
\centering

\caption{Age and grade for the pre-operative (BraTS 2020 [Tr], BraTS 2020 [V], BraTS 2020 [Te], and Phase 3 [Pre]; for brevity, we refer to the BraTS 2020 MRIs as BraTS in the column headers) and post-operative patient cohorts divided into the training and test sets, Phase 3 (Post, Tr) and Phase 3 (Post, Te), respectively. The number of patients is reported as $n$.}\label{tab:cohorts}
\setlength{\tabcolsep}{2.5pt}
\hspace*{-0.5cm}
\begin{tabular}{rcccccc}
\Xhline{2\arrayrulewidth}
\multicolumn{1}{c}{\textbf{Metric$\downarrow$}}                                                                                           & \begin{tabular}[c]{@{}c@{}}\textbf{BraTS}\\\textbf{(Tr)} \\    \textbf{n = 369}\end{tabular} & \begin{tabular}[c]{@{}c@{}}\textbf{BraTS}\\    \textbf{(V)} \\ \textbf{n   = 125}\end{tabular} & \begin{tabular}[c]{@{}c@{}}\textbf{BraTS}\\\textbf{(Te)}\\    \textbf{n   = 166}\end{tabular} & \begin{tabular}[c]{@{}c@{}}\textbf{Phase 3 (Pre)}\\    \textbf{n   = 100}\end{tabular} & \begin{tabular}[c]{@{}c@{}}\textbf{Phase 3} \\ \textbf{(Post, Tr)}\\    \textbf{n = 464}\end{tabular} & \begin{tabular}[c]{@{}c@{}}\textbf{Phase 3} \\    \textbf{(Post, Te)}\\   \textbf{ n   = 40}\end{tabular} \\
\hline
\begin{tabular}[c]{@{}r@{}}Mean   ± std. dev. \\ age and range \\ (min-max) (years)\end{tabular} & \begin{tabular}[c]{@{}c@{}}61±12\\    (19--87)\end{tabular}              & \begin{tabular}[c]{@{}c@{}}57±14   \\    (22--86)\end{tabular}           & \begin{tabular}[c]{@{}c@{}}61±12   \\    (18--87)\end{tabular}            & \begin{tabular}[c]{@{}c@{}}56±12\\  (21--79)\end{tabular}             & \begin{tabular}[c]{@{}c@{}}56±11\\ (18--84)\end{tabular}                    & \begin{tabular}[c]{@{}c@{}}55±10\\ (26--76)\end{tabular}                        \\
\hline
Sex   (female/male)                                                                              & Unknown                                                                 & Unknown                                                                 & Unknown                                                                  & 41/59                                                                & 152/312                                                                    & 11/29                                                                          \\
                     \hline                                                                            & \multicolumn{5}{c}{\textbf{Grade}}                                                                                                                                                                                                                                                                                                                                                        &                                                                                \\
                     \hline
WHO                                                                                              & Unknown                                                                 & \multirow{3}{*}{Unknown}                                                & \multirow{3}{*}{Unknown}                                                 & IV: 100                                                              & IV:   464                                                                  & IV:   40                                                                       \\
LGG                                                                                              & 76                                                                      &                                                                         &                                                                          & --–                                                                    & --–                                                                          & --–                                                                              \\
HGG                                                                                              & 293                                                                     &                                                                         &                                                                          & 100                                                                  & 464                                                                        & 40                                                                            \\
\Xhline{2\arrayrulewidth}
\end{tabular}
\end{table}

\begin{figure}[ht!]
    \centering
    \includegraphics[width=1\columnwidth]{
    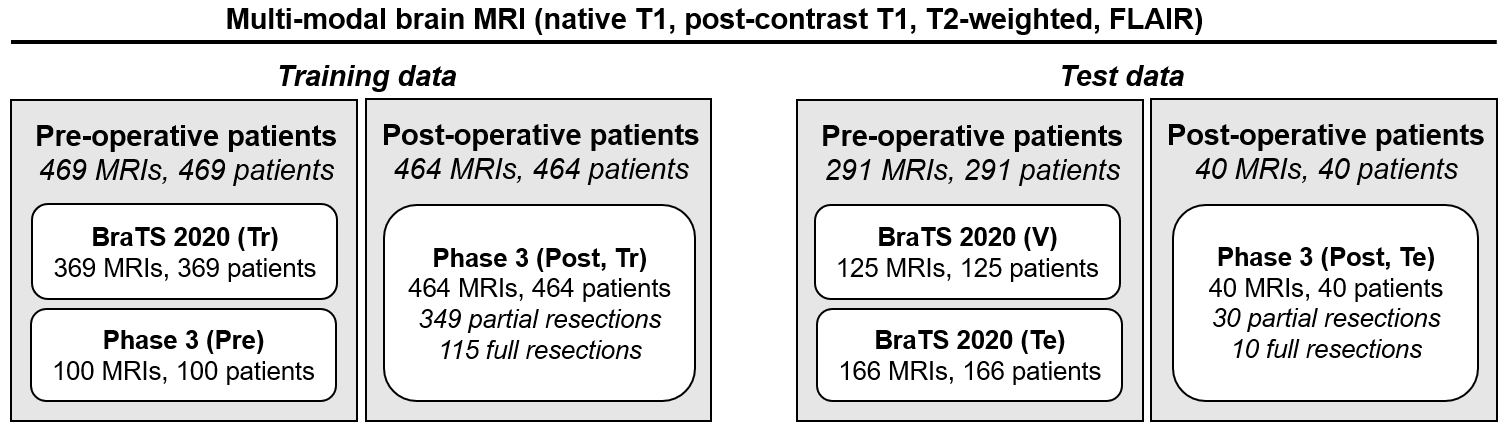}
    \caption{The datasets used in this study are divided into pre and post-operative MRIs, and into the training and test data.}
    \label{fig:dataset_split}
\end{figure}

The pre- and post-operative clinical phase 3 study cohorts (Phase 3 [Pre] and Phase 3 [Post], with 100 and 504 patients, respectively) were acquired in 92 sites (years 2009--2011) as part of a large pivotal clinical study in newly diagnosed GBM patients~\citep{Chinot2014ClinicalTrial}. As only MRI scans acquired prior to therapy (baseline scan) were considered in this study, each patient contributed to a single MRI dataset. For each patient, the MRI acquisition consisted of a) axial native T1 b) 3D axial, coronal, and sagittal post-Gadolinium T1, c) 2D axial T2-weighted fast spin echo, and d) 2D axial FLAIR, with  $\leq 5$~mm slice thickness, $\leq 1.5$~mm interslice gap, 0.1 mmol/kg body weight of the Gadolinium contrast. The post-operative MRIs were captured 1 to 57 days after the surgical intervention.

\begin{figure}[ht!]
    \centering
    \includegraphics[width=0.9\columnwidth]{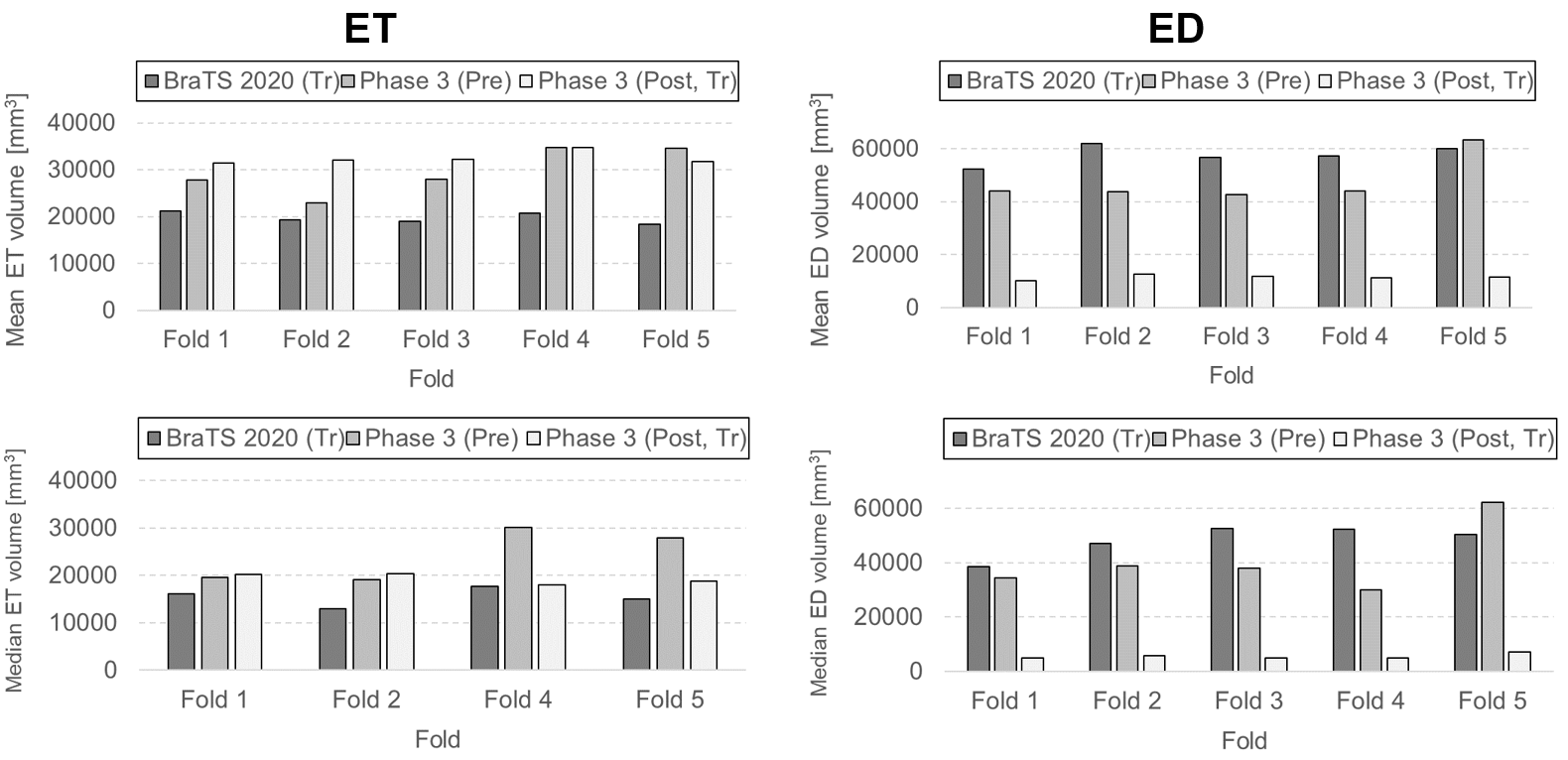}
    \caption{Data stratification across the folds in training data---we report the mean and median volume [mm$^3$] of ET and ED in each fold, and in each dataset.}
    \label{fig:fold_distribution}
\end{figure}

\begin{figure}[ht!]
    \centering
    \begin{tabular}{cc}
        \multicolumn{1}{c}{~~~~~~~~a)} & ~~~~~~~~~~~~~~~~~~~~~~b)\\
        \multicolumn{2}{c}{\includegraphics[width=1\columnwidth]{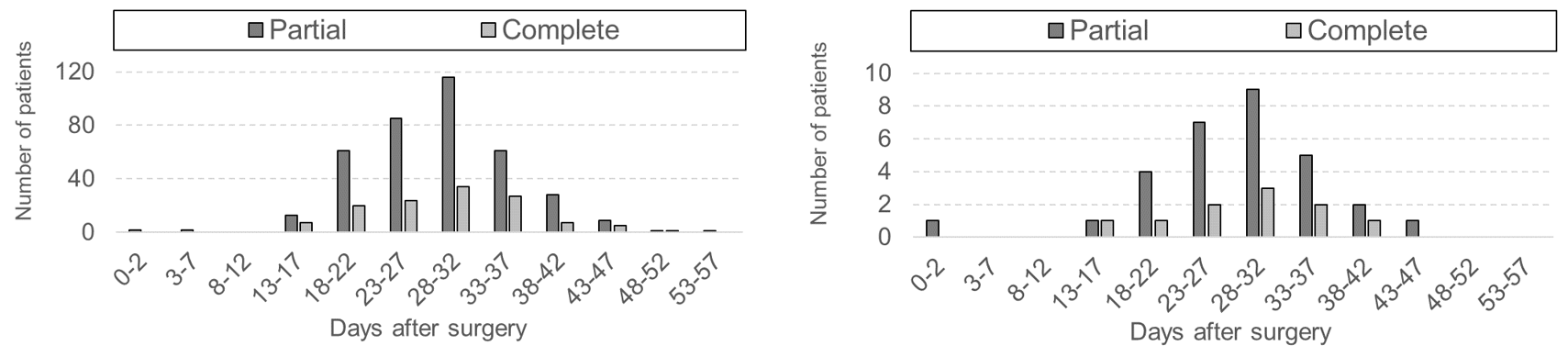}}
    \end{tabular}

    \caption{Distribution of the a)~training and b)~test patients from Phase 3 (Post, Tr) and Phase 3 (Post, Te), respectively, according to the number of days after surgery for intended partial and complete resections.}
    \label{fig:distribution_surgery}
\end{figure}

\begin{figure}[ht!]
    \centering
    \includegraphics[width=1\columnwidth]{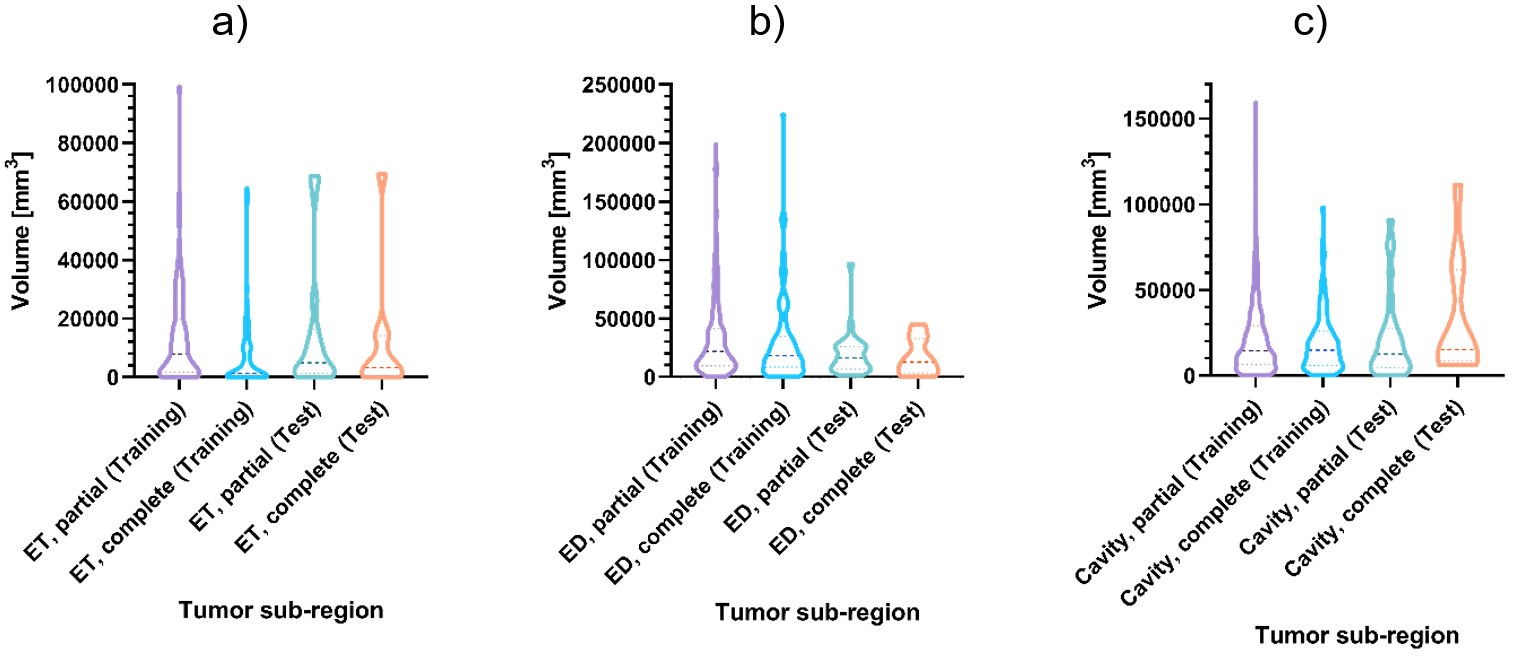}
    \caption{Distribution of a) ET, b) ED, and c) surgical cavity in training and test Phase 3 (Post) subsets, Phase 3 (Post, Tr) and Phase 3 (Post, Te), respectively.}
    \label{fig:tumor_subregion_distribution}
\end{figure}

The pre- and post-operative Phase 3 cohorts were both split into a training and a test set at the patient level (Figure~\ref{fig:dataset_split}), the former being used to train the deep learning models and the latter for quantifying the performance of the algorithms. The Phase 3 (Post) cohort was split into the training and test subsets (Phase 3 [Post, Tr] and Phase 3 [Post, Te]) which were stratified according to the distribution of partial and full resections (surgery status refers here to the purpose of the surgery and not to the actual extent of the resection), time of the image data acquisition, measured in days after the surgical intervention, and the volume of ET, ED, and cavity (Figures~\ref{fig:dataset_split}--\ref{fig:tumor_subregion_distribution}, Table~\ref{tab:descriptive_stats}). This stratification allows us to maintain similar characteristics of training and test sets, hence we do not bias the training/assessment of models toward specific (well-represented) tumors while omitting other (under-represented) tumors. The brain tumor segmentation training set included MRIs from 469 pre-operative and 464 post-operative patients whereas the test set consisted of images from 291 pre-operative subjects, together with 40 post-operative test MRIs, also utilized for validating the automated bidimensional calculation according to the RANO criteria.

\begin{table}[ht!]
\scriptsize
\centering

\setlength{\tabcolsep}{1.5pt}
\caption{Descriptive statistics of the training and test MRIs belonging to the Phase 3 (Post) set: Phase 3 (Post, Tr) and Phase 3 (Post, Te), respectively.}\label{tab:descriptive_stats}
\hspace*{-0.3cm}
\begin{tabular}{rrrcrrcrrcrrcrrcrrc}
\Xhline{2\arrayrulewidth}
\textbf{Class→}           & \multicolumn{5}{c}{\textbf{ET}}                      &               & \multicolumn{5}{c}{\textbf{ED}}                                     && \multicolumn{5}{c}{\textbf{Cavity}}                                 \\
\cline{2-6} \cline{8-12} \cline{14-18}
                 & \multicolumn{2}{c}{\textbf{Partial}} & &\multicolumn{2}{c}{\textbf{Complete}} & & \multicolumn{2}{c}{\textbf{Partial}} & &\multicolumn{2}{c}{\textbf{Complete}} & & \multicolumn{2}{c}{\textbf{Partial}} & &\multicolumn{2}{c}{\textbf{Complete}} \\
                 \cline{2-3} \cline{5-6} \cline{8-9} \cline{11-12} \cline{14-15} \cline{17-18}
\textbf{Metric↓}          & \textbf{Train.}        & \textbf{Test}   &     & \textbf{Train.}        & \textbf{Test} &        & \textbf{Train.}        & \textbf{Test}  &      & \textbf{Train.}        & \textbf{Test}  &       & \textbf{Train.}        & \textbf{Test}   &     & \textbf{Train.}        & \textbf{Test}         \\
\hline
Patients         & 349           & 30  &        & 115           & 10 &          & 349           & 30    &      & 115           & 10    &       & 349           & 30       &   & 115           & 10           \\
Min. {[}mm$^3${]}   & 0             & 0       &    & 0             & 0     &       & 0             & 1536   &     & 0             & 0   &         & 0             & 12      &    & 0             & 6181         \\
Mean {[}mm$^3${]}   & 13212         & 14567   &    & 6270          & 11152   &     & 33459         & 19346  &     & 29324         & 16943   &     & 21618         & 21825    &   & 19101         & 32494        \\
Median {[}mm$^3${]} & 7962          & 4971    &    & 1259          & 3324       &  & 21669         & 16249   &    & 18240         & 12813 &       & 14586         & 12695   &    & 14874         & 15290        \\
Max. {[}mm$^3${]}   & 99277         & 68692   &    & 64472         & 69413       & & 198806        & 96546      & & 224359        & 44846 &       & 159203        & 90837    &   & 98148         & 111230      \\
\Xhline{2\arrayrulewidth}
\end{tabular}
\end{table}

The clinical treatment trial was approved by the applicable independent ethics committees and institutional review boards~\citep{Chinot2014ClinicalTrial}. The Informed Consent Form permitted the use of study results for future medical research. The data has been anonymized for this purpose. The BraTS data comes from the public domain database, which can be utilized for non-commercial scientific use.

\subsection{Manual segmentation and bidimensional (RANO) measurements}\label{sec:manual_procedure}

As described by \cite{bakas2019identifying}, the BraTS pre-operative cohorts were segmented by 1--4 readers who followed the same manual analysis procedure, and all annotations were reviewed by the experienced neuro-radiologists. The delineation included several tumor sub-regions which in turn were used to obtain the enhancing tumor (ET) and the (edema) ED areas~\citep{Haller2013}.

\begin{figure}[ht!]
    \centering
    \hspace*{-2cm}
    \includegraphics[width=1\columnwidth]{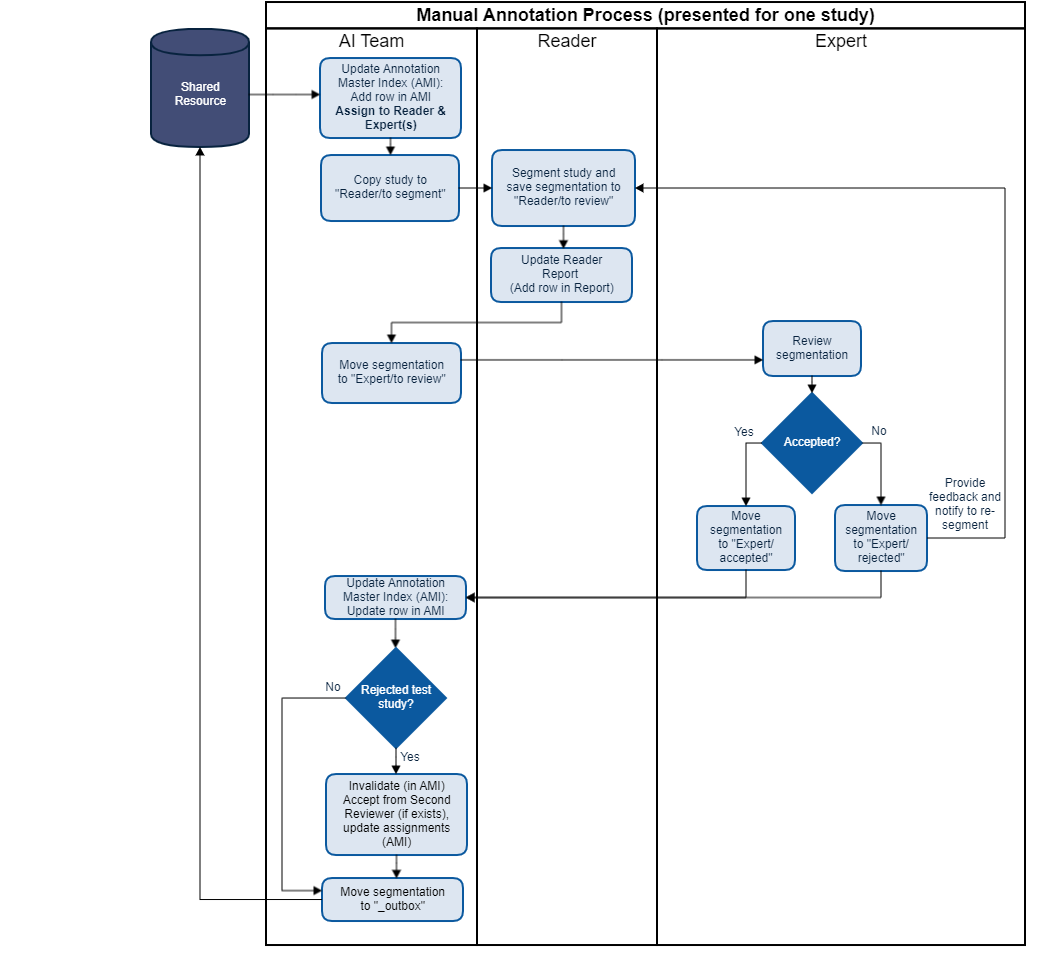}
    \caption{Manual segmentation procedure used to obtain delineations of brain tumors (alongside additional information related to the contouring, including the time of analysis and the confidence of a reader). The delineations are approved by senior radiologists. The raters provided their confidence for all GBM sub-regions, and the bidimensional measurements as per RANO were performed by each reader for Phase 3 (Post, Te). These measures were aggregated across all readers for all MRIs, and the average, weighted average (according to YOE), median, minimum, and maximum RANO was obtained for each patient as well.}
    \label{fig:annotation_process}
\end{figure}

For the pre- and post-operative phase 3 cohort, we enlisted two expert radiologists (Reader 1 and Reader 2, with 20 and 18 Years Of clinical Experience, YOE), together with five experienced raters (Readers 3--7 with 6, 6, 5, 3, and 2 YOE) who were tasked to analyze the input MRI scans---during the analysis process, the readers were always presented with all available MRI sequences, therefore they can benefit from all image data captured for each patient. The readers manually delineated ET, ED, and cavity (in post-surgery MRIs), following a rigorous annotation procedure (Figure~\ref{fig:annotation_process}) which expands on the BraTS annotation process by introducing the following steps:
\begin{itemize}
    \item We capture the readers’ confidences in the quality of their annotations.
    \item We incorporate a redrawing/improvement step.
    \item We obtain the manual bidimensional RANO measurements.
\end{itemize}
\noindent Here, the sub-regions were contoured by a single reader (Reader 3--7) and then approved subsequently by one of the expert radiologists (either Reader 1 or Reader 2), that, if necessary, highlighted the tumor components that needed to be redrawn (redrawing/improvement step). For the 40 test post-surgery patients (Phase 3 [Post, Te]) used to validate the automated bidimensional RANO measurements, the manual segmentation had to be approved by both expert radiologists (Reader 1 and Reader 2). All raters used the ITK-SNAP software (version 3.6.0), and we captured the time required to contour all tumor sub-regions by each reader. The raters provided additional information regarding their confidence for all GBM sub-regions, according to the following scale: 
\begin{enumerate}
    \item I am not confident at all (almost guessing).
    \item I am not confident and require advice from a senior reader.
    \item I am confident to some extent, but I would prefer to have it reviewed by a senior reader.
    \item I am fully confident.
\end{enumerate}
\noindent Finally, bidimensional measurements as per RANO were performed by each reader for Phase 3 (Post, Te). These measures were additionally aggregated across all readers for all MRIs, and the average, weighted average (according to YOE), median, minimum, and maximum RANO was obtained for each patient.

\subsection{Segmenting brain tumors from multi-modal MRI using deep learning}\label{sec:algorithm}

\begin{figure}[ht!]
    \centering
    \includegraphics[width=0.85\columnwidth]{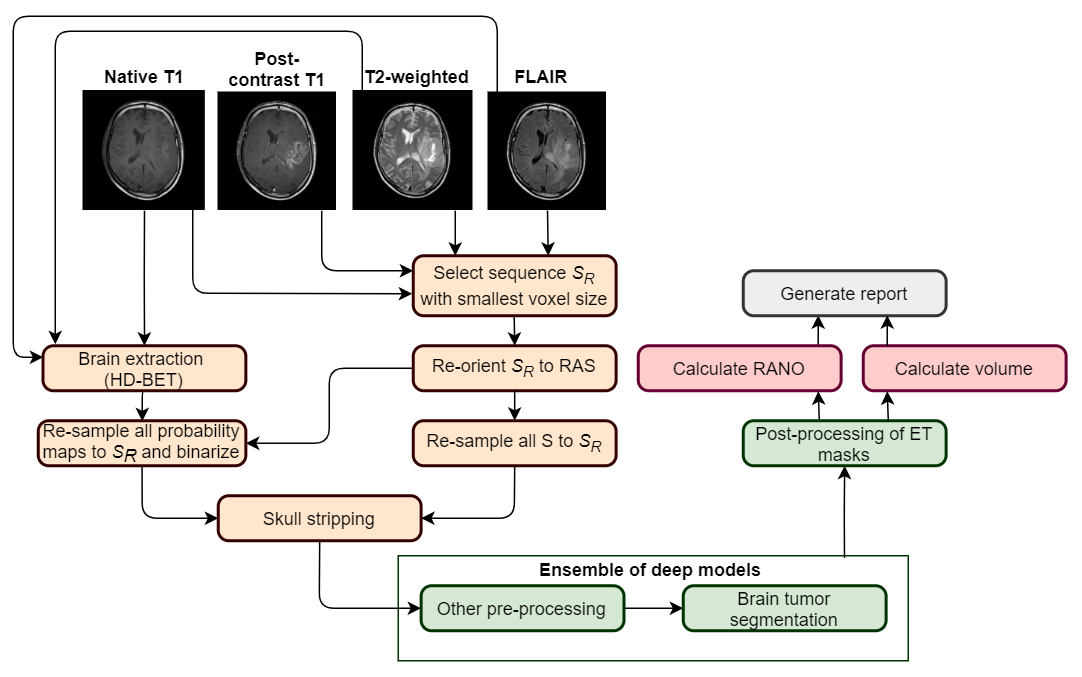}
    \caption{A schematic view of our processing pipeline. In light orange, we render the pre-processing steps performed before feeding the data into the ensemble model, in green---an additional pre-processing created using an nnU-Net engine for each model within our ensemble (including data augmentation and z-scoring), together with the segmentation step and post-processing, in red---calculation of bidimensional and volumetric tumor metrics, finally in gray---generating the final report, e.g., in a DICOM-RT format.}
    \label{fig:flowchart}
\end{figure}

Our segmentation algorithm operates on the co-registered native T1, post-contrast T1, T2-weighted and FLAIR MRI sequences, all in the axial orientation (Figure~\ref{fig:flowchart}). The approach is split into three pivotal steps: (\textit{i})~pre-processing, including re-orientation and re-sampling of sequences, together with skull stripping (brain extraction), (\textit{ii})~brain tumor segmentation using an ensemble of deep learning models, and (\textit{iii})~post-processing of the resulting segmentation map.

\subsubsection{Pre-processing}

\begin{figure}[ht!]
    \centering
    \includegraphics[width=1\columnwidth]{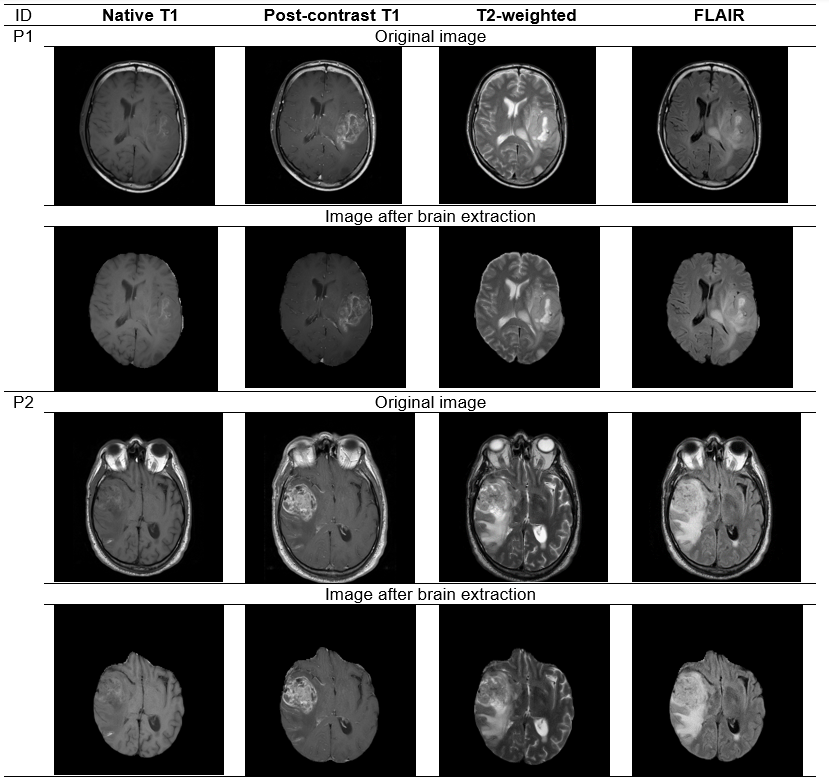}
    \caption{Two examples of post-operative MRIs from Phase 3 (Post, Tr) before and after brain extraction (skull stripping). Skull stripping is one of the pre-processing steps in the analysis pipeline. P1 and P2 denote patient 1 and patient 2, respectively.}
    \label{fig:skullstripping}
\end{figure}

In the pre-processing step, we use HD-BET~\citep{Isensee2019HDBET} for brain extraction in each sequence separately (see the examples in Figure~\ref{fig:skullstripping}). In parallel, the sequence with the smallest voxel size $S_R$ is determined and re-oriented to the Right, Anterior, Superior (RAS) coordinate system. The brain probability maps, alongside all other sequences $S$ are later linearly re-sampled to $S_R$. Finally, the brain probability maps are binarized, and skull stripping is performed in all sequences.

\subsubsection{Ensemble of the confidence-aware nnU-Nets}

A single ensemble of five confidence-aware nnU-Nets (see the detailed model architecture in the Supplementary Material) trained over different training sets is used to process both pre- and post-operative MRIs, and to automatically detect ET, ED, and cavity areas. It receives, as inputs, pre-processed native T1, post-contrast T1, T2-weighted and FLAIR sequences and performs z-score normalization, hence each sequence is normalized independently through extracting its mean intensity value and dividing by standard deviation. The base deep models are assembled by averaging softmax probabilities. During the prediction, the full MRIs with up to four modalities are inputted.

\newcommand{\LossFunction}{\mathcal{L}}
\newcommand{\Class}{\mathcal{C}}
\newcommand{\ConfidenceMultiplier}{\alpha}

The models are trained with the loss function $\LossFunction$, being the averaged cross-entropy and soft DICE, averaged across all target classes ET, ED, and Cavity: 
\begin{equation}
    \LossFunction=\frac{\LossFunction({\rm ET})+\LossFunction({\rm ED})+\LossFunction({\rm Cavity})}{3}.
\end{equation}

\noindent The soft DICE loss becomes

\begin{equation}
    \LossFunction_{\rm DICE}=1-\frac{2\left|P\cap GT\right|}{\left|P\right|^2+\left|GT\right|^2},
\end{equation}

\noindent where $P$ and $GT$ indicate the predicted and ground-truth (GT) segmentation masks. To exploit the reader’s confidence available for Phase 3 (Pre) and Phase 3 (Post, Tr) training MRIs, the loss was modified:

\begin{equation}
    \LossFunction'=\frac{\LossFunction'({\rm ET})+\LossFunction'({\rm ED})+\LossFunction'({\rm Cavity})}{3},
\end{equation}
\noindent where $\LossFunction'(\Class)=\LossFunction(\Class)\cdot \ConfidenceMultiplier(\Class)$, $\Class\in \{{\rm ET}, {\rm ED}, {\rm Cavity}\}$, and $\ConfidenceMultiplier$ denotes the multiplier related to the corresponding level of confidence. The following mapping is exploited: 0.5 for the confidence of 1 (i.e., ``I am not confident at all\dots''), 0.75 for 2, 1.25 for 3, and 1.5 for 4---the higher the confidence of the GT segmentation is, the larger impact on the loss it has. The mapping was non-linear to better separate acceptable (confidence 3 and 4) and non-acceptable (confidence 1 and 2) cases. For the MRIs (or specific sub-region types) without reported raters’ confidences, $\ConfidenceMultiplier({\rm ET})=\ConfidenceMultiplier({\rm ED})=\ConfidenceMultiplier({\rm Cavity})=1$ is used during training.

\subsubsection{Post-processing}

\begin{figure}[ht!]
    \centering
    \includegraphics[width=1\columnwidth]{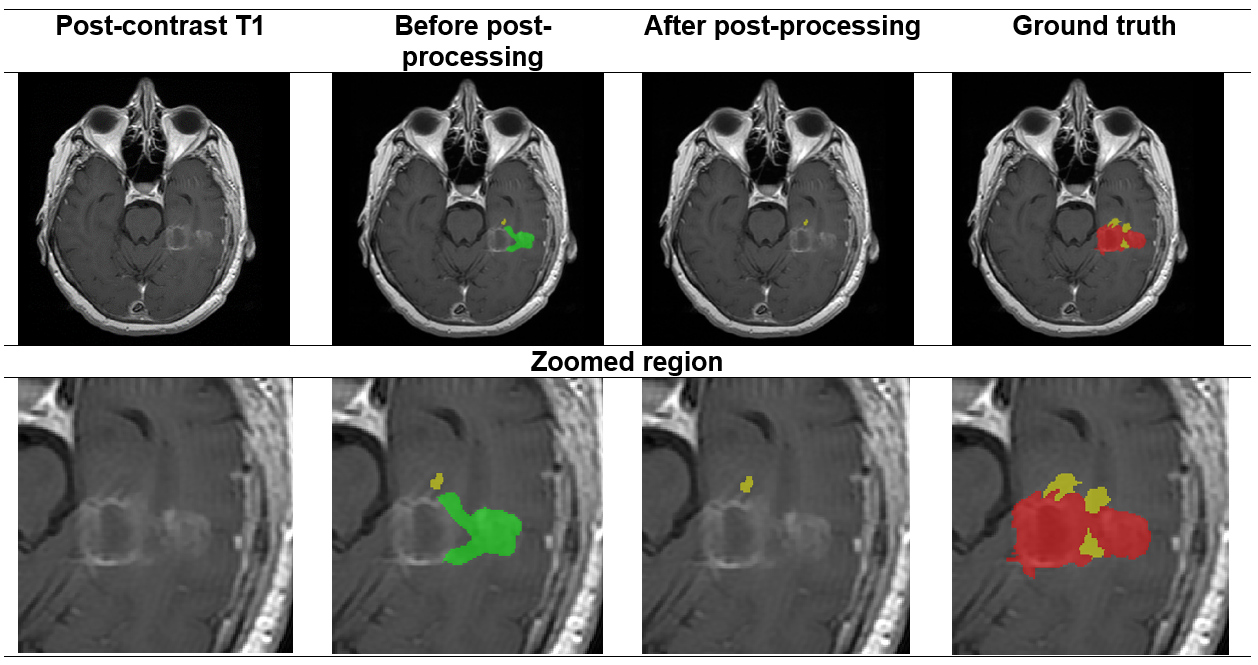}
    \caption{The proposed post-processing routine allows for pruning false positive ET sub-regions. Green shows the predicted ET region, yellow shows predicted ED, whereas red shows the surgical cavity in the ground truth. A part of the image is zoomed in the second row.}
    \label{fig:postprocessing_example}
\end{figure}

Most MRIs used in our study were performed more than 5 days after surgery. As a result, the MRI acquisition does not avoid the contrast enhancement associated with the surgical intervention. Therefore, since enhancing tumor area are typically located within invaded tissues, in order to avoid false positives (FPs) that may arise from hyper-intense regions in post-contrast T1 along the surgical cavity~\citep{2014Lescher}, as a post-processing step, we remove all detected ET voxels that were not neighboring to ED (in 3D)---an example segmentation before and after the suggested post-processing is rendered in Figure~\ref{fig:postprocessing_example}.

\subsubsection{Splitting training data into stratified folds and evaluating the model}

All 933 training MRIs were split into five non-overlapping folds used for training base models in our segmentation ensemble. To maintain the original distribution of ET and ED volumes within all subsets, each training set (BraTS 2020 [Tr], Phase 3 [Pre], and Phase 3 [Post, Tr]) was split into five stratified folds, according to the ET and ED volume distributions (Figure~\ref{fig:fold_distribution}). The corresponding folds from each set were combined (i.e., Fold 1 from BraTS 2020 [Tr], Phase 3 [Pre], and Phase 3 [Post, Tr], and so forth). Each base model in an ensemble is trained for 1000 epochs using stochastic gradient descent with Nesterov momentum ($\mu=0.99$) on a training set composed of four different folds, with one fold kept aside and acting as the validation set. The batch included two patches of size 208×238×196, and 250 batches were processed within an epoch. Training-time data augmentation encompassed random patch scaling within (0.7, 1.4), random rotation, random gamma correction within (0.7, 1.5), and random mirroring~\citep{10.3389/fncom.2019.00083}.

\subsubsection{Automated bidimensional measurements}

We introduce a fully automatic algorithm for calculating bidimensional measurements in post-contrast T1 sequences, strictly following the current RANO criteria~\citep{Ellingson2017}. For each detected ET region in the input 3D volume, the algorithm exhaustively searches for the longest segment (major diameter) over all slices, and then for the corresponding longest perpendicular diameter, with the tolerance of 5 degrees inclusive. Such segments are valid if they (\textit{i})~are fully inscribed in ET, and (\textit{ii})~are both at least 10 mm long (otherwise, the lesion is not measurable). Finally, the product of the perpendicular diameters is calculated. If there are more measurable ET regions, the sum of up to five largest products is returned. We refer to this algorithm as Automated RANO (Diameters). In addition, we introduce an alternative and more robust version of the automated two-dimensional RANO (Automated RANO [Product]), that exhaustively optimizes the product of diameters instead of the maximum diameter. We found this approach to be less sensitive to small alterations in the contour of the lesions.

\subsubsection{Quantitative and statistical analysis}

To evaluate the manual and automated segmentation, we computed the DICE coefficient, the Jaccard’s index (also referred to as the Intersection over Union, IoU), sensitivity and specificity\footnote{Note that for BraTS 2020~(V) and BraTS 2020~(Te) we are unable to report the IoU scores, as they are not calculated by the validation server (the ground-truth annotations are not publicly available).}. For those parameters, the larger value obtained the better, with 1.0 denoting a perfect score. The DICE coefficient is calculated as:
\begin{equation}
  {\rm DICE(P,GT)}=\frac{2 \cdot \left|{\rm P}\cap {\rm GT}\right|}{\left|{\rm P}\right|+\left|{\rm GT}\right|}=\frac{\rm 2\cdot TP}{\rm 2 \cdot TP + FP + FN},
  \end{equation}
\noindent whereas for IoU we have:
\begin{equation}
  {\rm IoU(P,GT)}=\frac{\left|{\rm P}\cap {\rm GT}\right|}{\left|{\rm P}\cup {\rm GT}\right|}=\frac{\rm TP}{\rm TP + FP + FN},
  \end{equation}
where P and GT are two segmentations (predicted and ground truth), and TP, FP, and FN are the numbers of true positives, false positives, and false negatives. Both DICE and IoU are the overlap metrics, but can notice that IoU penalizes single instances of wrong segmentation more than DICE, therefore IoU tends to quantify the ``worst'' case average performance of the segmentation algorithm.

In addition, the 95$^{\rm th}$ percentile of Hausdorff distance (H95; the smaller, the better), which quantifies the contours' similarity, was also calculated~\citep{bakas2019identifying}. Since the shape of the ET contours may easily affect the RANO calculation, e.g., jagged contours could result in over-pessimistic bidimensional measurements, investigating both overlap measures (e.g., DICE/IoU) together with H95 is pivotal to thoroughly quantify the algorithm's capabilities, as it should simultaneously obtain maximum overlap metrics and maintain minimum distance between the automatic and manual contours. The inter-rater and algorithm-rater agreement for bidimensional and volume measurements was evaluated using the Intraclass Correlation Coefficient (ICC) calculated on a single measurement, absolute-agreement, two-way random-effects model. The R package IRR (Inter Rater Reliability, version 0.84.1) was used for ICC, whereas GraphPad Prism 9.1.2 for calculating the Spearman’s correlation coefficient ($\SpearmanCoefficient$) and all other statistics.

\subsubsection{Implementation details and code availability}

The proposed deep learning pipeline was implemented in Python 3.7 with the PyTorch 1.6.0 backend. Our segmentation model is built upon an established open-sourced nnU-Net framework~\citep{Isensee2021nnunet} available at \url{https://github.com/MIC-DKFZ/nnUNet}, whereas for brain extraction we utilize the HD-BET software~\citep{Isensee2019HDBET} available at \url{https://github.com/MIC-DKFZ/HD-BET}. To ensure full reproducibility of the deep learning segmentation algorithm, we present the details of the deep model architecture in the Supplementary Material.

\section{Results}\label{sec:results}

This section gathers the results of our experimental study. We focus on the pivotal aspects of the pipeline, including the quality of tumor segmentation (Section~\ref{sec:BT_Segmentation}), RANO measurements (Section~\ref{sec:rano_results}), correlations between RANO and volumetric measurements (Section~\ref{sec:rano_volume}), and its processing time (Section~\ref{sec:proc_time}).

\subsection{Segmentation of brain tumors}\label{sec:BT_Segmentation}

The performance of our segmentation algorithm was evaluated for a pre-surgery set using BraTS 2020 (V) (which included 125 patients). For computing the performance matrix in the post-surgery setting, for ET we used 32 patients with existing ground-truth ET regions from Phase 3 (Post, Te), whereas for evaluating ED, this was 39 patients with existing ground-truth ED regions. 

For pre-operative patients, mean DICE for ET was 0.744 (95\% CI: 0.690--0.799) with median DICE of 0.871 (25\% percentile--75\% percentile, 25p--75p: 0.776--0.917). The corresponding mean H95 was 39.624 mm with median H95 of 2.000 mm (25p--75p: 1.000--3.399 mm). The cavity was erroneously detected in pre-operative patients: 8/125 patients (6.4\%) for BraTS 2020 (V) and 10/166 patients (6.0\%) for BraTS 2020 (Te), respectively. The cavity detection results are gathered in Table~\ref{tab:cavity_for_preoperative}, in which we confront our model with the vanilla nnU-Nets trained on either post-operative training MRIs, or all training MRIs.

% The performance parameters could not be derived for ED using the BraTS dataset. 

\begin{table}[ht!]
\scriptsize
\caption{The surgical cavity detection results obtained over pre-operative patients (any detections are false positives) for BraTS 2020~(V) and BraTS 2020~(Te) by the vanilla nnU-Net trained on post-operative training MRIs (Post) or all training MRIs, and by the proposed model. The best results are boldfaced, and the worst results are underlined.}\label{tab:cavity_for_preoperative}
\begin{tabular}{rccc}
\Xhline{2\arrayrulewidth}
\textbf{Metric$\downarrow$}                             & \textbf{nnU-Net (Post)}                 & \textbf{nnU-Net}    & { \textbf{Proposed}}  \\
\hline
\rowcolor[HTML]{FFFFFF} 
                                             \multicolumn{4}{c}{\cellcolor[HTML]{FFFFFF}\textbf{BraTS   2020 (V), 125 pre-operative patients}}  \\
                                            \hline
\rowcolor[HTML]{FFFFFF} 
Number of patients with detected cavity     & \uline{102}         & { \textbf{8}}         & { \textbf{8}}                                   \\
\rowcolor[HTML]{FFFFFF} 
Percentage of patients with detected cavity & \uline{81.60\%}     & { \textbf{6.40\%}}    & {\textbf{ 6.40\%}}                              \\
\rowcolor[HTML]{FFFFFF} 
Mean vol. of detected cavity [mm$^3$]          & \uline{10468.03}    & { \textbf{5611.75}}   & 5931.38                                   \\
\rowcolor[HTML]{FFFFFF} 
Min. vol. of detected cavity [mm$^3$]          & \textbf{1}                          & 4               & \uline{6}                                         \\
\rowcolor[HTML]{FFFFFF} 
Max. vol. of detected cavity [mm$^3$]          & \uline{70927}       & \textbf{22576}           & 25491                                     \\
\rowcolor[HTML]{FFFFFF} 
Median vol. of detected cavity [mm$^3$]        & \uline{5019}        & { \textbf{792}}       & 981                                       \\
\hline
\rowcolor[HTML]{FFFFFF} 
                                             \multicolumn{4}{c}{\cellcolor[HTML]{FFFFFF}\textbf{BraTS   2020 (Te), 166 pre-operative patients}} \\
                                            \hline
\rowcolor[HTML]{FFFFFF} 
Number of patients with detected cavity     & \uline{128}         & 12              & { \textbf{10}}                                  \\
\rowcolor[HTML]{FFFFFF} 
Percentage of patients with detected cavity & \uline{77.10\%}     & 7.20\%          & { \textbf{6.00\%}}                              \\
\rowcolor[HTML]{FFFFFF} 
Mean vol. of detected cavity [mm$^3$]          & \uline{9772.15}     & { 735.17}    & \textbf{468.1}                            \\
\rowcolor[HTML]{FFFFFF} 
Min. vol. of detected cavity [mm$^3$]          & { \textbf{2}}                             & 3               & \uline{34}                \\
\rowcolor[HTML]{FFFFFF} 
Max. vol. of detected cavity [mm$^3$]          & \uline{85582}       & { 6904}      & \textbf{2565}                             \\
\rowcolor[HTML]{FFFFFF} 
Median vol. of detected cavity [mm$^3$]        & \uline{4404}        & \textbf{73}              & 202                                      \\
\Xhline{2\arrayrulewidth}
\end{tabular}
\end{table}

For post-operative patients, the median DICE for ET was 0.735 (25p--75p: 0.588--0.801). The mean DICE was 0.692 (95\% CI: 0.628--0.757), 0.677 (0.631--0.724) and 0.691 (0.604--0.778)  for ET, ED, and surgical cavity. The respective values for mean H95 were 9.221 mm (6.437--12.000 mm), 9.455 mm (7.176--11.730 mm) and 7.956 mm (5.938--9.975 mm). The mean DICE for ET was significantly larger for the data segmented with the highest confidence by the readers, 0.749 (95\% CI: 0.698--0.800) for confidence level 4 compared to 0.599 (95\% CI: 0.452--0.746) for the remainder (reader confidence level 1 to 3). 

\begin{figure}[ht!]
    \centering
    \includegraphics[width=1\columnwidth]{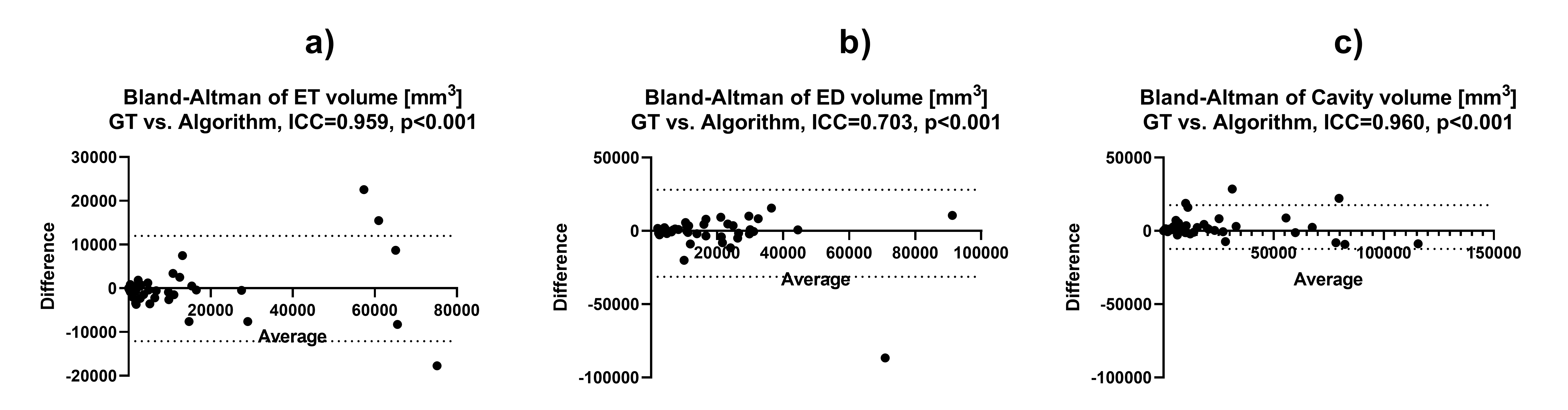}
    \caption{Bland-Altman plots for a) ET, b) ED, and c) cavity volumes extracted manually (GT) and the algorithm for all post-operative MRIs showing that the model is in almost perfect agreement for ET and surgical cavity.}
    \label{fig:bland_altman_et_ed_cavity}
\end{figure}

Exploiting the readers’ confidence improved the segmentation capabilities of the deep learning models, significantly outperforming the other techniques as indicated with DICE/H95 analysis. The mean and median values of all quality metrics were improved for the ET area, as well as for T2/FLAIR abnormalities (ED). Finally, detecting the surgical cavity automatically resulted in further improvements in the quality of ET delineation\footnote{For detailed segmentation performance plots with and without the utilization of the readers' confidence, see the Supplementary Material.}. The automated volumetric measurements (in mm$^3$) for ET and the surgical cavity sub-regions were in almost perfect agreement with the ground-truth segmentations (ICC: 0.959, $p<0.001$, Figure~\ref{fig:bland_altman_et_ed_cavity}a, ICC: 0.960, $p<0.001$, Figure~\ref{fig:bland_altman_et_ed_cavity}c, respectively), whereas for ED the agreement was ICC: 0.703 ($p<0.703$; Figure~\ref{fig:bland_altman_et_ed_cavity}b).

\begin{table}[ht!]
\scriptsize
\setlength{\tabcolsep}{3pt}
\caption{The results obtained in post-surgery setting, with all MRI sequences included, as well as if T2-weighted or FLAIR are voluntarily omitted. The best results are boldfaced, whereas the worst---underlined.}\label{tab:removing_modalities}
\setlength{\tabcolsep}{2.5pt}
\centering
\hspace*{-0.5cm}
\scalebox{0.9}{
\begin{tabular}{rrrrrrrrrrrrrrrrrrrrrr}
\Xhline{2\arrayrulewidth}
        &                         & \textbf{Without}                    & \textbf{Without}  &                &                         & \textbf{Without}                    & \textbf{Without}     &             &                        & \textbf{Without}                    & \textbf{Without}    &              \\
\textbf{Metric$\downarrow$}       &           \textbf{All MRI}              &  \textbf{T2w}                    &  \textbf{FLAIR}       &           &          \textbf{All MRI}          &  \textbf{T2w}                    &  \textbf{FLAIR}       &           &               \textbf{All MRI}         &  \textbf{T2w}                    &  \textbf{FLAIR}                  \\
        \hline
              \multicolumn{13}{c}{\textbf{DICE}}                                                                                                                                                                                                                                                                               \\
              \hline
              & \multicolumn{3}{c}{\textbf{ET}}                                                                           && \multicolumn{3}{c}{\textbf{ED}}                                                                           && \multicolumn{3}{c}{\textbf{Cavity}}                                                                       \\
              \cline{2-4} \cline{6-8} \cline{10-12}
25p           & 0.588                          & \textbf{0.617}                          & \uline{0.099}  && \textbf{0.615}                      &    0.529                          & \uline{0.156}  && \textbf{0.596}                          & \uline{0.227}  & 0.488                          \\
Median        & 0.735                          &\textbf{ 0.741}                          & \uline{0.674}  && \textbf{0.708}                          & 0.674                          & \uline{0.250}  && 0.774                          & \uline{0.680}  & \textbf{0.786}                          \\
75p           & \textbf{0.801}                          & 0.795                          & \uline{0.790}  && \textbf{0.769}                          & 0.751                          & \uline{0.470}  && \textbf{0.908}                          & \uline{0.877}  & 0.883                          \\
Mean          & \textbf{0.692}                          & 0.684                          & \uline{0.521}  && \textbf{0.677}                          & 0.654                          & \uline{0.302}  && \textbf{0.691}                          & \uline{0.570}  & 0.641                          \\
Lower 95\% CI & \textbf{0.628}                          & 0.613                          & \uline{0.401}  && \textbf{0.631}                          & 0.606                          & \uline{0.228}  && \textbf{0.604}                          & \uline{0.463}  & 0.542                          \\
Upper 95\% CI & \textbf{0.757}                          & 0.755                          & \uline{0.641}  && \textbf{0.724}                          & 0.702                          & \uline{0.375}  && \textbf{0.778}                          & \uline{0.678}  & 0.741                          \\
\hline
              \multicolumn{13}{c}{\textbf{IoU}}\\
              \hline
              & \multicolumn{3}{c}{\textbf{ET}}                                                                           && \multicolumn{3}{c}{\textbf{ED}}                                                                           && \multicolumn{3}{c}{\textbf{Cavity}}                                                                         \\
              \cline{2-4} \cline{6-8} \cline{10-12}
25p           & 0.416                          & \textbf{0.447}                          & \uline{0.062}  && \textbf{0.444}                          & 0.360                          & \uline{0.085}  && \textbf{0.425}                          & \uline{0.130}  & 0.323                          \\
Median        & 0.581                          & \textbf{0.588}                          & \uline{0.509}  && \textbf{0.548}                          & 0.509                          & \uline{0.143}  && 0.631                          & \uline{0.517}  & \textbf{0.648}                          \\
75p           & \textbf{0.668}                          & 0.660                          & \uline{0.653}  &&\textbf{ 0.625}                          & 0.601                          & \uline{0.307}  && \textbf{0.832}                          & \uline{0.781}  & 0.791                          \\
Mean          & \textbf{0.553}                         & 0.547                          & \uline{0.413}  && \textbf{0.528}                          & 0.502                          & \uline{0.200}  && \textbf{0.583}                          & \uline{0.469}  & 0.538                          \\
Lower 95\% CI & \textbf{0.488}                          & 0.478                          & \uline{0.311}  && \textbf{0.478}                          & 0.451                          & \uline{0.143}  && \textbf{0.495}                          & \uline{0.369}  & 0.442                          \\
Upper 95\% CI & \textbf{0.617}                          & \textbf{0.617}                          & \uline{0.515}  && \textbf{0.578}                          & 0.554                          & \uline{0.258}  && \textbf{0.671}                          & \uline{0.569}  & 0.633                          \\
              \hline
              \multicolumn{13}{c}{\textbf{H95}}                                                                                                                                                                                                                                                                                \\
              \hline
              & \multicolumn{3}{c}{\textbf{ET}}                                                                           && \multicolumn{3}{c}{\textbf{ED}}                                                                           && \multicolumn{3}{c}{\textbf{Cavity}}                                                                        \\
              \cline{2-4} \cline{6-8} \cline{10-12}
25p           & \textbf{3.400}                          & 3.518                          & \uline{4.894}  && 5.315                          &\textbf{ 4.599}                          & \uline{15.120} && \textbf{3.241}                          & \uline{4.486}  & 3.871                          \\
Median        & 8.204                          & \textbf{7.620}                          & \uline{8.279}  && \textbf{8.348}                          & 8.634                          & \uline{20.780} && \textbf{6.391}                          & \uline{9.575}  & 6.529                          \\
75p           & 12.610                         & \textbf{12.330}                        & \uline{20.700} && \textbf{12.120}                         & 12.340                         & \uline{34.250} && \textbf{11.150}                         & \uline{20.010} & 11.870                         \\
Mean          & \textbf{9.221}                          & 9.532                          & \uline{14.060} && \textbf{9.455}                          & 10.010                         & \uline{24.960} && \textbf{7.956}                          & \uline{13.560} & 10.650                         \\
Lower 95\% CI & \textbf{6.437}                          & 6.663                          & \uline{8.524}  && \textbf{7.176}                          & 7.712                          & \uline{20.280} && \textbf{5.938}                          & \uline{9.334}  & 6.945                          \\
Upper 95\% CI & \textbf{12.000}                         & 12.400                         & \uline{19.590} && \textbf{11.730}                         & 12.310                         & \uline{29.650} && \textbf{9.975}                          & \uline{17.780} & 14.360                         \\
\hline
              \multicolumn{13}{c}{\textbf{Sensitivity}}                                                                                                                                                                                                                                                                        \\
              \hline
              & \multicolumn{3}{c}{\textbf{ET}}                                                                           && \multicolumn{3}{c}{\textbf{ED}}                                                                           && \multicolumn{3}{c}{\textbf{Cavity}}                                                                        \\
              \cline{2-4} \cline{6-8} \cline{10-12}
25p           & 0.626                          & \textbf{0.643}                          & \uline{0.081}  &&\textbf{ 0.628}                          & 0.543                          & \uline{0.085}  && \textbf{0.491}                          & \uline{0.133}  & 0.348                          \\
Median        & 0.752                          & \textbf{0.768}                          & \uline{0.628}  && \textbf{0.728}                          & 0.705                          & \uline{0.491}  && \textbf{0.756}                          & \uline{0.572}  & 0.746                          \\
75p           & 0.872                          & \textbf{0.880}                          & \uline{0.783}  && \textbf{0.775}                          & 0.762                          & \uline{0.324}  && \textbf{0.925}                          & \uline{0.844}  & 0.898                          \\
Mean          & \textbf{0.720}                          & \textbf{0.720}                          & \uline{0.504}  && \textbf{0.695}                          & 0.665                          & \uline{0.213}  && \textbf{0.662}                          & \uline{0.522}  & 0.613                          \\
Lower 95\% CI & \textbf{0.646}                          & 0.644                          & \uline{0.381}  && \textbf{0.651}                          & 0.613                          & \uline{0.148}  && \textbf{0.563}                          & \uline{0.410}  & 0.506                          \\
Upper 95\% CI & 0.794                          &\textbf{ 0.796}                          & \uline{0.627}  && \textbf{0.738}                          & 0.717                          & \uline{0.277}  && \textbf{0.762}                          & \uline{0.633}  & 0.719                          \\
\hline
              \multicolumn{13}{c}{\textbf{Specificity}}                                                                                                                                                                                                                                                                        \\
              \hline
             & \multicolumn{3}{c}{\textbf{ET}}                                                                           && \multicolumn{3}{c}{\textbf{ED}}                                                                           && \multicolumn{3}{c}{\textbf{Cavity}}                                                                          \\
             \cline{2-4} \cline{6-8} \cline{10-12}
25p           & 0.9995                         & \uline{0.9992} & \textbf{0.9997}                         && 0.9992                         & \uline{0.9991} & \textbf{0.9999}                        && 0.9996                         & \textbf{0.9997}                         & \uline{0.9995} \\
Median        & \textbf{0.9998}                         & \textbf{0.9998}                         & \textbf{0.9998}                         && \uline{0.9996} & \uline{0.9996} & \textbf{1.0000}                         && \uline{0.9998} & \textbf{0.9999}                         & \uline{0.9998} \\
75p           & \uline{0.9999} & \uline{0.9999} & \textbf{1.0000}                         && \uline{0.9998} & \uline{0.9998} & \textbf{1.0000}                        && \uline{0.9999} & \textbf{1.0000}                         & \textbf{1.0000}                         \\
Mean          & 0.9995                         & \uline{0.9994} & \textbf{0.9997}                         && 0.9992                         & \uline{0.9991} & \textbf{0.9999}                         && \uline{0.9997} & \textbf{0.9998}                         & \uline{0.9997} \\
Lower 95\% CI & 0.9993                         & \uline{0.9991} & \textbf{0.9995}                         && \uline{0.9986} & \uline{0.9986} & \textbf{0.9999}                         && \uline{0.9995} & \textbf{0.9997}                         & \uline{0.9995} \\
Upper 95\% CI & 0.9998                         & \uline{0.9997} & \textbf{0.9999}                         && \uline{0.9997} & \uline{0.9997} & \textbf{1.0000}                         && \uline{0.9998} & \textbf{0.9999}                         & \uline{0.9998}\\
\Xhline{2\arrayrulewidth}
\end{tabular}}
\end{table}

The algorithm is trained and originally designed to work with a set of four MRI sequences: T1 pre/post contrast, T2-weighted and FLAIR. To check the robustness of our segmentation technique in the case of missing sequences, we voluntarily omitted either T2-weighted or FLAIR during the prediction process (Table~\ref{tab:removing_modalities}). Here, we did not remove native T1 and post-contrast T1, as those are pivotal in visualizing and quantifying ET~\citep{10.1093/neuonc/nou221}. After removing T2-weighted, the mean (median) DICE scores changed by ↓0.009 (↑0.006) for ET, ↓0.024 (↓0.033) for ED, and ↓0.121 (↓0.094) for cavity. Omitting FLAIR resulted in changing the mean (median) DICE by ↓0.172 (↓0.061) for ET, ↓0.376 (↓0.458) for ED, and ↓0.050 (↑0.012) for cavity. The differences in the overlap metrics (DICE and IoU) are statistically significant for ET and ED only in the case of removing FLAIR ($p<0.0001$, Wilcoxon matched-pairs signed rank test). Similarly, automated RANO (Diameters) and Automated RANO (Product) are statistically significantly different after removing FLAIR ($p<0.0001$), but not after removing the T2-weighted sequence. For the surgical cavity, however, the decrease in these metrics is significant if either T2-weighted or FLAIR are missing ($p<0.0001$).

\begin{table}[ht!]
\scriptsize
\setlength{\tabcolsep}{1.5pt}
\caption{The Intraclass Correlation Coefficient (ICC) obtained for the bidimensional measurements (RANO) across all readers and aggregations of readers’ responses, as well as for Automated RANO---two algorithms, one maximizing the major and perpendicular diameters sequentially (referred as Diameters), and the other optimizing the product of the perpendicular diameters (Product). Best results in green, worse in red.}\label{tab:ICC_readers}
\begin{tabular}{rrrrr}
\Xhline{2\arrayrulewidth}
                                      & &  & \multicolumn{1}{r}{\textbf{Auto. RANO}} & \multicolumn{1}{r}{\textbf{Auto. RANO}} \\
                                                                            & \multicolumn{1}{c}{\textbf{GT (Diam.)}} & \multicolumn{1}{c}{\textbf{GT (Prod.)}} & \multicolumn{1}{r}{\textbf{(Diameters)}} & \multicolumn{1}{r}{\textbf{(Product)}} \\
                                                                            \hline
\textbf{Reader 1}                     & \cellcolor[HTML]{BEE3CA}0.760               & \cellcolor[HTML]{D7EDDF}0.701             & \cellcolor[HTML]{DFF0E6}0.681                             & \cellcolor[HTML]{E3F2E9}0.671                           \\
\textbf{Reader 2}                     & \cellcolor[HTML]{DEF0E5}0.683               & \cellcolor[HTML]{A5D9B4}0.822             & \cellcolor[HTML]{93D2A4}0.866                             & \cellcolor[HTML]{96D3A7}0.858                           \\
\textbf{Reader 3}                     & \cellcolor[HTML]{F9B0B2}0.406               & \cellcolor[HTML]{F99395}0.328             & \cellcolor[HTML]{F8888A}0.299                             & \cellcolor[HTML]{F88587}0.292                           \\
\textbf{Reader 4}                     & \cellcolor[HTML]{F99496}0.332               & \cellcolor[HTML]{FABFC2}0.447             & \cellcolor[HTML]{FBF2F5}0.583                             & \cellcolor[HTML]{FBE1E3}0.537                           \\
\textbf{Reader 5}                     & \cellcolor[HTML]{ECF6F1}0.650               & \cellcolor[HTML]{FBD9DC}0.516             & \cellcolor[HTML]{FACFD1}0.489                             & \cellcolor[HTML]{FAC4C7}0.460                           \\
\textbf{Reader 6}                     & \cellcolor[HTML]{FBF4F7}0.588               & \cellcolor[HTML]{FAC4C7}0.461             & \cellcolor[HTML]{FABDBF}0.441                             & \cellcolor[HTML]{FAB3B5}0.414                           \\
\textbf{Reader 7}                     & \cellcolor[HTML]{FBD8DB}0.515               & \cellcolor[HTML]{F9AEB0}0.401             & \cellcolor[HTML]{F9ABAE}0.394                             & \cellcolor[HTML]{F9A3A6}0.373                           \\
\textbf{Average}                      & \cellcolor[HTML]{DCEFE3}0.688               & \cellcolor[HTML]{F5F9F9}0.628             & \cellcolor[HTML]{E9F4EE}0.657                             & \cellcolor[HTML]{F9FBFD}0.617                           \\
\textbf{Weighted   average}           & \cellcolor[HTML]{B7E0C4}0.777               & \cellcolor[HTML]{BAE1C6}0.771             & \cellcolor[HTML]{ABDBB9}0.807                             & \cellcolor[HTML]{B9E1C6}0.772                           \\
\textbf{Median}                       & \cellcolor[HTML]{FBF9FB}0.601               & \cellcolor[HTML]{FAD5D7}0.505             & \cellcolor[HTML]{FAD2D5}0.499                             & \cellcolor[HTML]{FAC8CB}0.472                           \\
\textbf{Minimum}                      & \cellcolor[HTML]{F88386}0.287               & \cellcolor[HTML]{F87274}0.240             & \cellcolor[HTML]{F86B6D}0.221                             & \cellcolor[HTML]{F8696B}0.215                           \\
\textbf{Maximum}                      & \cellcolor[HTML]{CDE9D6}0.725               & \cellcolor[HTML]{92D1A4}0.868             & \cellcolor[HTML]{7FC993}0.915                             & \cellcolor[HTML]{7DC991}0.919                           \\
\textbf{GT (Diameters)}               & \cellcolor[HTML]{FFFFFF}                    & \cellcolor[HTML]{76C68C}0.935             & \cellcolor[HTML]{99D4A9}0.852                             & \cellcolor[HTML]{9CD6AC}0.843                           \\
\textbf{GT (Product)}                 & \multicolumn{1}{r}{}                        & \multicolumn{1}{r}{}                      & \cellcolor[HTML]{77C68C}0.934                             & \cellcolor[HTML]{73C589}0.944                           \\
\textbf{Auto. RANO (Diameters)} & \multicolumn{1}{r}{}                        & \multicolumn{1}{r}{}                      & \cellcolor[HTML]{FFFFFF}                                  & \cellcolor[HTML]{63BE7B}0.981                \\
\Xhline{2\arrayrulewidth}
\end{tabular}
\end{table}

\subsection{Inter-rater agreement for RANO bidimensional measurements}\label{sec:rano_results}

\begin{figure}[ht!]
    \centering
    \includegraphics[width=0.63\columnwidth]{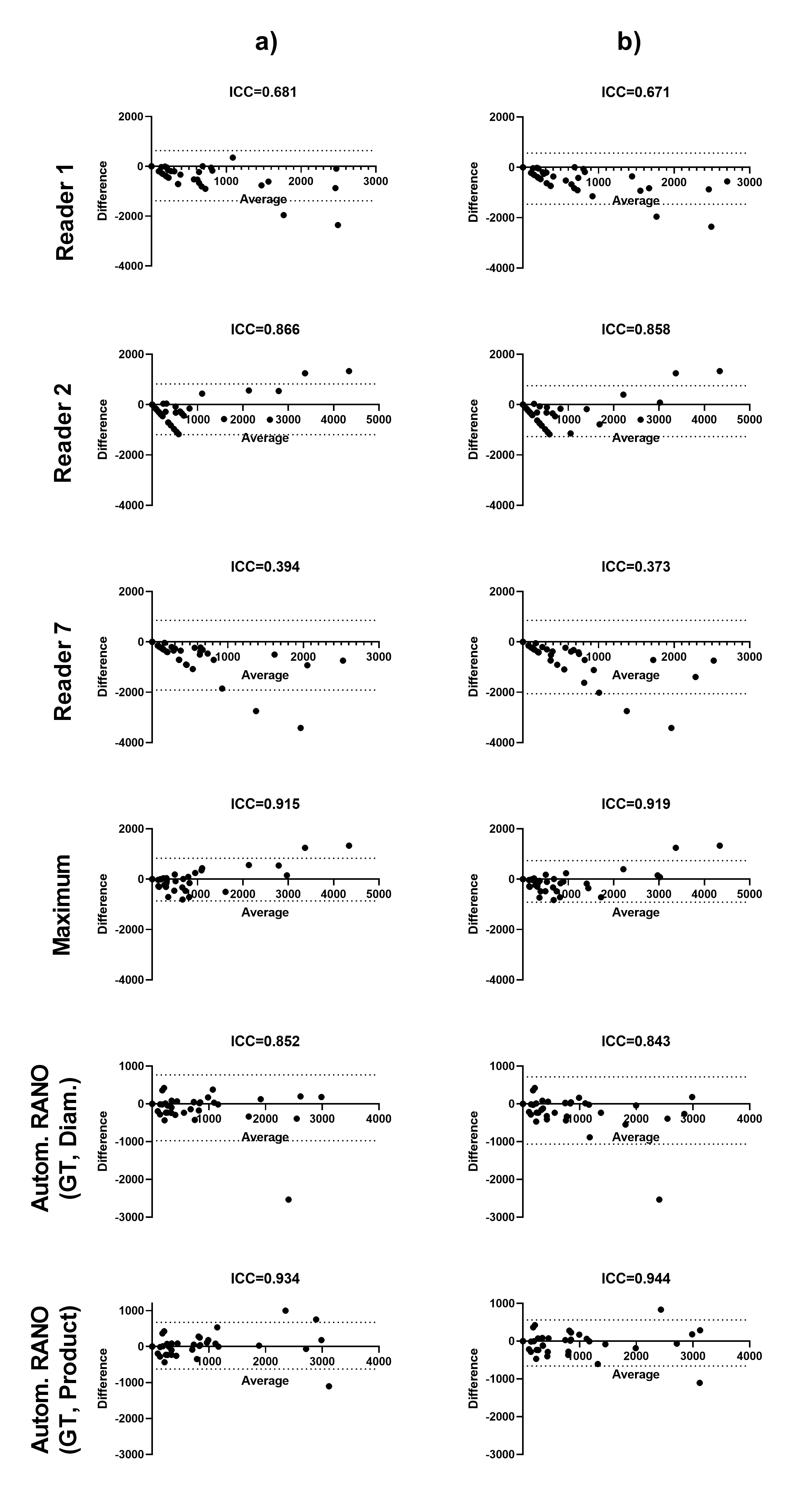}
    \caption{Bland-Altman plots for a) Automated RANO (Diameters) and b) Automated RANO (Product) (both calculated using the algorithm derived ET segmentation using all post-operative MRIs) and \textit{i})~senior radiologists (Reader 1 and Reader 2), \textit{ii})~the least experienced rater (Reader 7, 2 YOE), \textit{iii})~the maximum RANO calculated for each patient across all the readers, \textit{iv})~Automated RANO (Diameters) calculated over the GT segmentations, and \textit{v})~Automated RANO (Product) calculated over the GT segmentations. Our algorithm consistently delivers the largest agreement with the senior radiologists and tends to return larger RANO as it carefully optimizes the RANO diameters.}
    \label{fig:bland_altman_auto_rano}
\end{figure}

There was a significant inter-rater variability and disagreement across the human readers for the bidimensional RANO calculation in the post-surgery setting (Phase 3 [Post, Te] dataset), with ICC: 0.220--0.960 (Table~\ref{tab:ICC_readers}). The agreement between the manual RANO provided by the raters and Automated RANO (Diameters) calculated over the predicted ET regions was ICC: 0.299--0.866 ($p<0.001$), whereas for Automated RANO (Product) it amounted to 0.292--0.858 ($p<0.001$). Automated RANO (Diameters) and Automated RANO (Product) were in strong agreement with the maximum values aggregated across all human raters, ICC: 0.915 ($p<0.001$) and 0.919 ($p<0.001$). Both versions of the automated RANO were consistently resulting in larger ICC with the most senior radiologists (Figure~\ref{fig:bland_altman_auto_rano}) compared to less-experienced readers (Table~\ref{tab:ICC_RANO}).

\begin{table}[ht!]
\scriptsize
\setlength{\tabcolsep}{2.3pt}
\caption{The Intraclass Correlation Coefficient (ICC) obtained for the bidimensional measurements (RANO) across all readers (R1--R7 in the column headers) and all aggregations of readers’ responses (Phase 3 [Post, Te] test MRIs). Best results in green, worse in red.}\label{tab:ICC_RANO}
\begin{tabular}{rrrrrrrrrrrr}
\Xhline{2\arrayrulewidth}
\multicolumn{1}{l}{}   & \multicolumn{1}{c}{\textbf{R2}} & \multicolumn{1}{c}{\textbf{R3}} & \multicolumn{1}{c}{\textbf{R4}} & \multicolumn{1}{c}{\textbf{R5}} & \multicolumn{1}{c}{\textbf{R6}} & \multicolumn{1}{c}{\textbf{R7}} & \multicolumn{1}{c}{\textbf{Avg.}} & \multicolumn{1}{c}{\textbf{W avg.}} & \multicolumn{1}{c}{\textbf{Median}} & \multicolumn{1}{c}{\textbf{Min.}} & \multicolumn{1}{c}{\textbf{Max.}} \\
\hline
\textbf{Reader 1}      & \cellcolor[HTML]{FBE0E3}0.580         & \cellcolor[HTML]{E8F4EE}0.711         & \cellcolor[HTML]{FBDCDE}0.566         & \cellcolor[HTML]{ABDBB9}0.829         & \cellcolor[HTML]{CCE9D6}0.765         & \cellcolor[HTML]{DEF0E5}0.730         & \cellcolor[HTML]{74C58A}0.934        & \cellcolor[HTML]{7BC890}0.920                 & \cellcolor[HTML]{90D1A2}0.880       & \cellcolor[HTML]{FAD2D5}0.534        & \cellcolor[HTML]{FBE6E9}0.601        \\
\textbf{Reader 2}      & \cellcolor[HTML]{FFFFFF}              & \cellcolor[HTML]{F87375}0.220         & \cellcolor[HTML]{FBECEF}0.621         & \cellcolor[HTML]{F99DA0}0.360         & \cellcolor[HTML]{F98E90}0.310         & \cellcolor[HTML]{F88183}0.265         & \cellcolor[HTML]{FBE1E4}0.585        & \cellcolor[HTML]{C6E7D1}0.776                 & \cellcolor[HTML]{F9ABAE}0.406       & \cellcolor[HTML]{F86D6F}0.200        & \cellcolor[HTML]{63BE7B}0.966        \\
\textbf{Reader 3}      & \cellcolor[HTML]{FFFFFF}              & \cellcolor[HTML]{FFFFFF}              & \cellcolor[HTML]{F99598}0.334         & \cellcolor[HTML]{C7E7D2}0.774         & \cellcolor[HTML]{DFF0E6}0.729         & \cellcolor[HTML]{BEE3C9}0.793         & \cellcolor[HTML]{D6EDDE}0.746        & \cellcolor[HTML]{FBDBDE}0.564                 & \cellcolor[HTML]{9DD6AD}0.856       & \cellcolor[HTML]{B3DFC0}0.814        & \cellcolor[HTML]{F87A7C}0.243        \\
\textbf{Reader 4}      & \cellcolor[HTML]{FFFFFF}              & \cellcolor[HTML]{FFFFFF}              & \cellcolor[HTML]{FFFFFF}              & \cellcolor[HTML]{F99698}0.335         & \cellcolor[HTML]{F99598}0.334         & \cellcolor[HTML]{F99093}0.317         & \cellcolor[HTML]{F6FAFA}0.684        & \cellcolor[HTML]{D9EEE1}0.741                 & \cellcolor[HTML]{FACBCE}0.512       & \cellcolor[HTML]{F99093}0.317        & \cellcolor[HTML]{FBDEE1}0.574        \\
\textbf{Reader 5}      & \cellcolor[HTML]{FFFFFF}              & \cellcolor[HTML]{FFFFFF}              & \cellcolor[HTML]{FFFFFF}              & \cellcolor[HTML]{FFFFFF}              & \cellcolor[HTML]{67C07E}0.960         & \cellcolor[HTML]{8CCF9E}0.889         & \cellcolor[HTML]{99D4AA}0.863        & \cellcolor[HTML]{E4F2EA}0.720                 & \cellcolor[HTML]{73C589}0.936       & \cellcolor[HTML]{FBE3E6}0.592        & \cellcolor[HTML]{F9AAAD}0.403        \\
\textbf{Reader 6}      & \cellcolor[HTML]{FFFFFF}              & \cellcolor[HTML]{FFFFFF}              & \cellcolor[HTML]{FFFFFF}              & \cellcolor[HTML]{FFFFFF}              & \cellcolor[HTML]{FFFFFF}              & \cellcolor[HTML]{86CD9A}0.899         & \cellcolor[HTML]{AFDDBD}0.821        & \cellcolor[HTML]{FBF8FB}0.660                 & \cellcolor[HTML]{79C78E}0.924       & \cellcolor[HTML]{FBE6E9}0.602        & \cellcolor[HTML]{F9989B}0.343        \\
\textbf{Reader 7}      & \cellcolor[HTML]{FFFFFF}              & \cellcolor[HTML]{FFFFFF}              & \cellcolor[HTML]{FFFFFF}              & \cellcolor[HTML]{FFFFFF}              & \cellcolor[HTML]{FFFFFF}              & \cellcolor[HTML]{FFFFFF}              & \cellcolor[HTML]{B6E0C3}0.807        & \cellcolor[HTML]{FBEAED}0.614                 & \cellcolor[HTML]{7BC890}0.920       & \cellcolor[HTML]{F0F7F5}0.696        & \cellcolor[HTML]{F8888A}0.288        \\
\textbf{Average}       & \cellcolor[HTML]{FFFFFF}              & \cellcolor[HTML]{FFFFFF}              & \cellcolor[HTML]{FFFFFF}              & \cellcolor[HTML]{FFFFFF}              & \cellcolor[HTML]{FFFFFF}              & \cellcolor[HTML]{FFFFFF}              & \cellcolor[HTML]{FFFFFF}             & \cellcolor[HTML]{73C589}0.937                 & \cellcolor[HTML]{6FC386}0.943       & \cellcolor[HTML]{FBEAED}0.615        & \cellcolor[HTML]{FBDFE2}0.578        \\
\textbf{Weighted avg.} & \cellcolor[HTML]{FFFFFF}              & \cellcolor[HTML]{FFFFFF}              & \cellcolor[HTML]{FFFFFF}              & \cellcolor[HTML]{FFFFFF}              & \cellcolor[HTML]{FFFFFF}              & \cellcolor[HTML]{FFFFFF}              & \cellcolor[HTML]{FFFFFF}             & \cellcolor[HTML]{FFFFFF}                      & \cellcolor[HTML]{BEE3C9}0.793       & \cellcolor[HTML]{FABBBE}0.459        & \cellcolor[HTML]{D0EAD9}0.758        \\
\textbf{Median}        & \cellcolor[HTML]{FFFFFF}              & \cellcolor[HTML]{FFFFFF}              & \cellcolor[HTML]{FFFFFF}              & \cellcolor[HTML]{FFFFFF}              & \cellcolor[HTML]{FFFFFF}              & \cellcolor[HTML]{FFFFFF}              & \cellcolor[HTML]{FFFFFF}             & \cellcolor[HTML]{FFFFFF}                      & \cellcolor[HTML]{FFFFFF}            & \cellcolor[HTML]{E1F1E8}0.725        & \cellcolor[HTML]{F9ADB0}0.413        \\
\textbf{Minimum}       & \cellcolor[HTML]{FFFFFF}              & \cellcolor[HTML]{FFFFFF}              & \cellcolor[HTML]{FFFFFF}              & \cellcolor[HTML]{FFFFFF}              & \cellcolor[HTML]{FFFFFF}              & \cellcolor[HTML]{FFFFFF}              & \cellcolor[HTML]{FFFFFF}             & \cellcolor[HTML]{FFFFFF}                      & \cellcolor[HTML]{FFFFFF}            & \cellcolor[HTML]{FFFFFF}             & \cellcolor[HTML]{F8696B}0.185       \\
\Xhline{2\arrayrulewidth}
\end{tabular}
\end{table}

\subsection{Correlation between bidimensional RANO and volumetric measurements}\label{sec:rano_volume}

\begin{figure}[ht!]
    \centering
    \includegraphics[width=1\columnwidth]{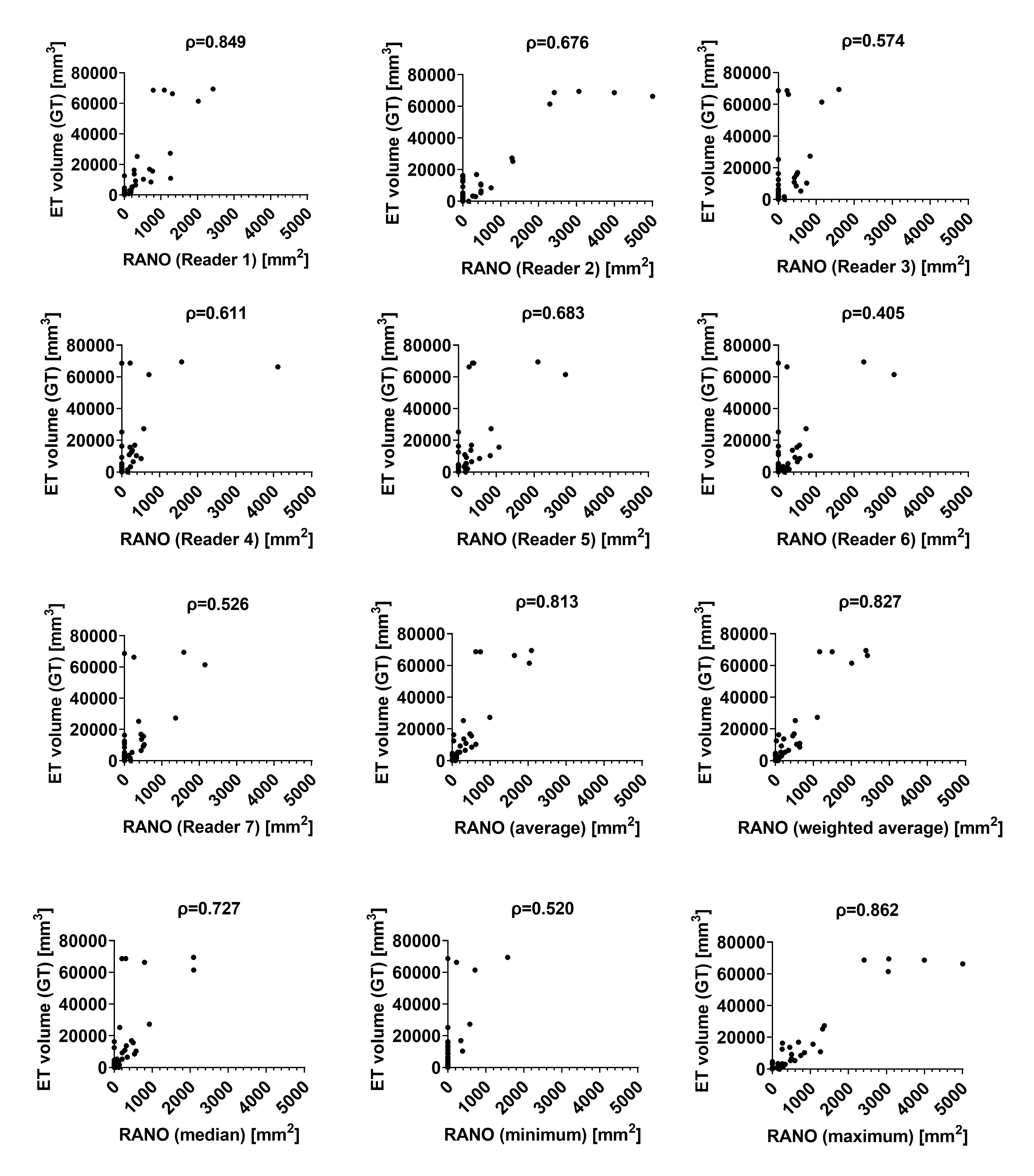}
    \caption{Correlation between the ground-truth ET volume [mm$^3$] and bidimensional measurements (RANO) [mm$^2$] calculated by the readers, and by aggregating the results delivered by the raters (average, weighted average according to the years of experience, median, minimum, and maximum across all readers for each patient), for all post-operative MRIs.}
    \label{fig:rano_correlation}
\end{figure}

\begin{figure}[ht!]
    \centering
    \includegraphics[width=0.8\columnwidth]{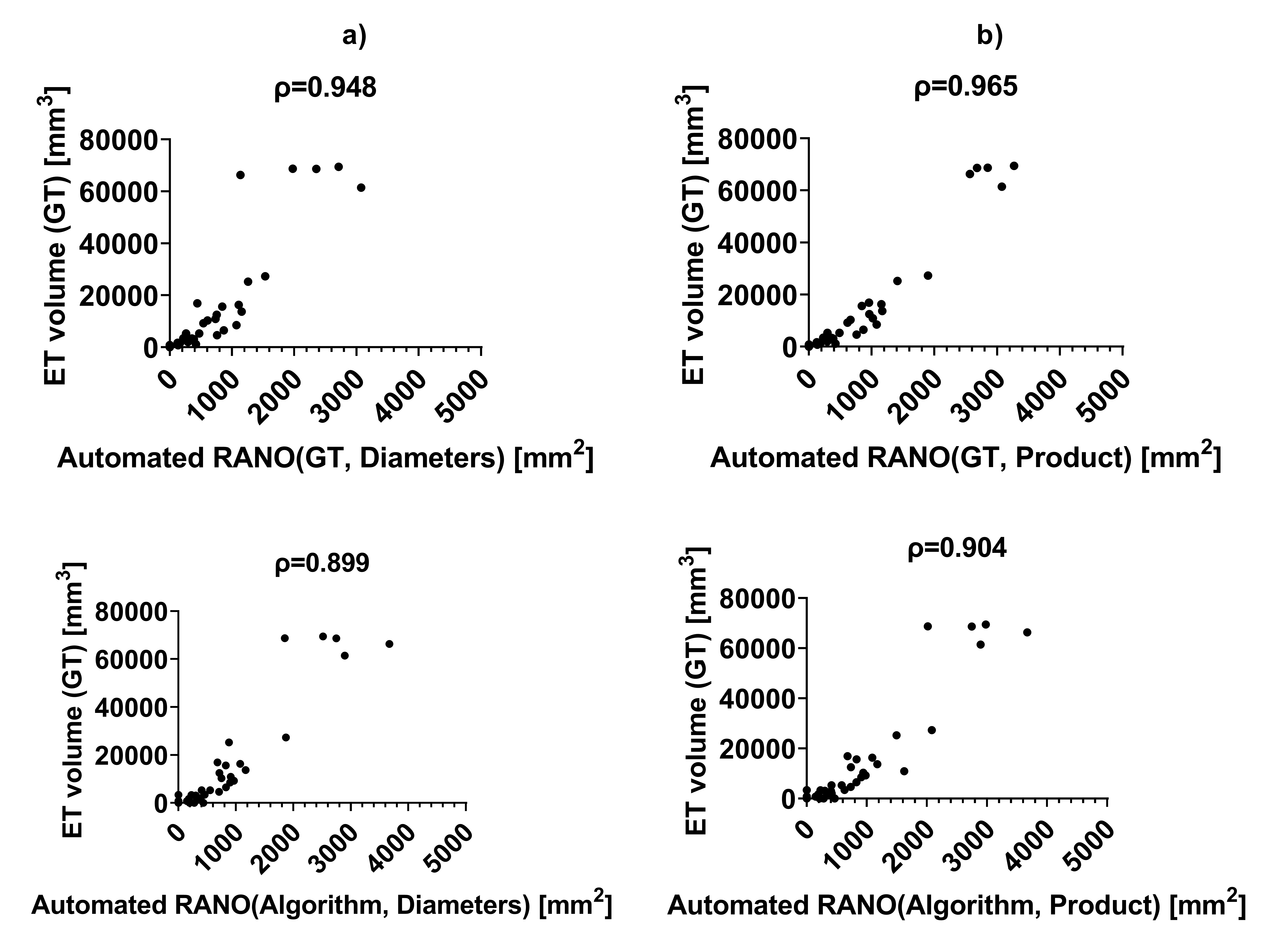}
    \caption{Correlation between the ground-truth ET volume [mm$^3$] and bidimensional measurements (RANO) [mm$^2$] obtained using a) Automated RANO (Diameters) and b) Automated RANO (Product) over the ground-truth (GT) segmentations and delivered by the algorithm for all post-operative MRIs. The RANO calculated using the algorithms is in almost perfect correlation with the ET volume, and the strength of this correlation is significantly larger than for RANO measured by the readers.}
    \label{fig:ET_RANO_correlation}
\end{figure}

The Spearman’s correlation coefficient $\SpearmanCoefficient$ between the RANO bidimensional values and ground-truth ET volumes calculated by the readers ranged from 0.405 (Reader 6) to 0.849 (Reader 1). For aggregated RANO values, $\SpearmanCoefficient$ was between 0.520 (minimum RANO across the raters) and 0.862 (maximum RANO across the raters), as presented in Figure~\ref{fig:rano_correlation}. Automated RANO strongly correlates with the ET volume (Figure~\ref{fig:ET_RANO_correlation}), $\SpearmanCoefficient$: 0.948 and 0.965 for GT (Automated RANO [Diameters] and Automated RANO [Product]), and $\SpearmanCoefficient$: 0.899 and 0.904 for our predictions (Automated RANO [Diameters] and Automated RANO [Product]).

\subsection{Processing time analysis}\label{sec:proc_time}

The experiments were executed on a high-performance computer equipped with an NVIDIA Tesla V100 GPU (32 GB) and 6 Intel Xeon E5-2680 (2.50 GHz) CPUs. The average end-to-end analysis time amounted to 148~s (with Automated RANO [Diameters]) and 829~s (with Automated RANO [Product]). This time includes brain extraction (81~s on average), other pre-processing routines (13~s), brain tumor segmentation (43~s), post-processing (1~s), and Automated RANO (Diameters) and Automated RANO (Product) calculation, 8~s and 678~s, respectively. In comparison, the average manual tumor segmentation time was 33, 39, 41, 50, and 36~mins for Readers 3--7. The algorithm is therefore at least 20.3× (Automated RANO [Diameters]) and 3.6× (Automated RANO [Product]) faster than the radiologists (we did not capture the time required for reader RANO calculation, only the duration of the manual image segmentation).

\section{Discussion}

Evaluating response to therapies in GBM depends extensively on the longitudinal radiological assessment of MRI scans to estimate change in tumor burden. This is based on two dimensional diameter measurements of enhancing tumors as well as on the qualitative estimation of T2/FLAIR abnormalities. Given the irregular shape and heterogeneous appearance of GBM lesions, this procedure is notoriously complex to perform and subject to high intra- and inter-reader variability, which, in turn, limits our ability to detect a therapeutic benefit in clinical trials and capture early patient response or progression in clinical practice. Volumetric analysis of all the lesions has also long been recognized as a potential alternative response endpoint to bidimensional RANO assessment. Because of the prohibitively highly tedious and time-consuming process of segmenting tumor lesions manually, integration of volumetric measurement into the clinical workflow is, however, only feasible with the advent of automation.

\begin{figure}[ht!]
    \centering
    \includegraphics[width=1\columnwidth]{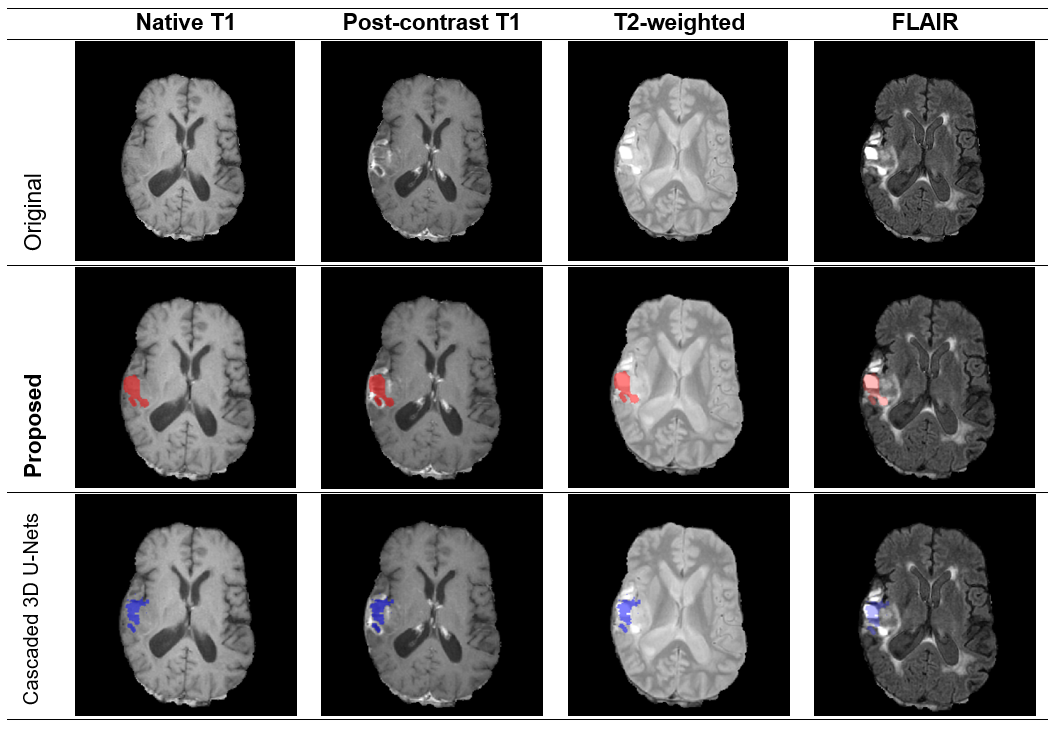}
    \caption{Incorrectly predicted cavity regions (in red) in pre-operative patients. This corresponded to the necrotic and non-enhancing tumor core (in blue, as returned by the cascaded 3D U-Nets~\citep{DBLP:conf/brainles-ws/KotowskiAMMZN20}) and therefore did not impact the RANO calculation.}
    \label{fig:cavity_segmentation}
\end{figure}

Our deep learning model is able to segment several tumor sub-regions simultaneously, including enhancing tumor area (ET) and T2/FLAIR abnormalities (ED). It has been developed by rigorously verifying 34 deep learning architectures (for the details of the investigated deep learning models, and the experimental results obtained for each of them, see the Supplementary Material) trained and tested on a large dataset consisting of 464 post-operative training and 40 test clinical multi-modal MRIs obtained from 92 institutions and multiple scanner types around the world that were manually annotated by 7 expert readers. Expanding the training sets to include additional pre-operative (469) patients as well as exploiting the readers’ confidence brought significant improvements of the segmentation quality metrics. Detecting the surgical cavity automatically allowed the model to further improve by capturing contextual information about the tumor’s sub-regions more comprehensively. This was achieved due to low prevalence of false positive and erroneous detection of surgical cavity and by the fact that when cavity was incorrectly annotated, this corresponded to the necrotic and non-enhancing tumor core with no impact on ET, ED and RANO tumor measurements (Figure~\ref{fig:cavity_segmentation}).

In the pre-surgery setting, the developed segmentation algorithm outperformed the recently introduced cascaded 3D U-Nets~\citep{DBLP:conf/brainles-ws/KotowskiAMMZN20} in all segmentation quality metrics\footnote{Computed using the independent validation server at https://ipp.cbica.upenn.edu/} (Table~\ref{tab:comparison_unet}). For post-operative patients, we also obtained high performance, comparable to the results reported by \cite{Chang2019GBM} and showing that the segmentation algorithm can be used successfully for pre- or post-operative patients. Although we are aware that confronting the results of the algorithms over different test sets may easily be misleading\footnote{However, we can anticipate that our segmentation model would likely outperform the one proposed by \cite{Chang2019GBM}, or at least work on par with that over the very same data---as already mentioned, the techniques utilized in our work allowed us to obtain Top-8 scores in the BraTS 2021 Challenge (out of 1600 participants).}, a large number of MRIs taken for investigation should compensate it and ``protect'' us from inferring over-optimistic conclusions with respect to the state of the art. 

We tested the robustness of the algorithm by voluntarily removing one imaging sequence (T2-weighted or FLAIR) from the model inputs. Overall, contrary to FLAIR and with the exception of the surgical cavity, the quality of the sub-region segmentation (ET and ED) as well as RANO bidimensional measurements are not statistically significantly compromised if T2-weighted images are not used (Table~\ref{tab:removing_modalities}). This reflects the preference of radiologists for using FLAIR for image evaluation as the nulling of the cerebrospinal fluid signal in this sequence augments the conspicuity of tumor lesions compared to T2-weighted sequences. Overall, removing T2-weighted from the input does not significantly affect ET or ED segmentation---the algorithm is sufficiently robust to be used with only native T1, post-contrast T1 and FLAIR (without T2-weighted sequences). We also evaluated the model performance early after surgery using the only patient dataset acquired within a few days after surgical intervention. Early post-operative scans are known to avoid surgically induced contrast enhancement, minimizing interpretative difficulties. While additional assessments would be beneficial in that setting, for the patient scanned 1 day after surgery the DICE and H95 value for ET (0.717, 5.99 mm) were comparable to the complete dataset (mean: 0.692, 9.221 mm, median: 0.735, 8.204 mm).

\begin{table}[ht!]
\scriptsize
\centering
\caption{The segmentation results obtained for BraTS 2020(V) using cascaded 3D U-Nets~\citep{DBLP:conf/brainles-ws/KotowskiAMMZN20} (with and without exploiting the expert knowledge, as proposed by~\cite{DBLP:conf/brainles-ws/KotowskiAMMZN20}), and the models introduced in this study. The best results for each metric are bold, and the worst---underlined.}\label{tab:comparison_unet}
\begin{tabular}{rrrrrrr}
\Xhline{2\arrayrulewidth}
        &     & \textbf{Cascaded}    & \textbf{Cascaded 3D} & \textbf{nnU-Net}            & \textbf{}    & { \textbf{}} \\
\multicolumn{2}{r}{\textbf{Metric$\downarrow$}}             & \textbf{3D U-Nets}    & \textbf{U-Nets with EK} & \textbf{(Post)}            & \textbf{nnU-Net}    & { \textbf{Proposed}} \\
\hline
        \multirow{5}{*}{\rotatebox[origin=c]{90}{\textbf{DICE}}}                               & Mean   & \uline{0.685}  & 0.694                               & 0.720                           & \textbf{0.750}            & 0.744                                    \\
                                       & s      & \uline{0.315}  & 0.310                                & \textbf{0.278}                 & 0.301           & 0.307                                    \\
                                       & Median & 0.839                          & 0.842                               & \uline{0.838}  & \textbf{0.872}           & 0.871                                    \\
                                       & 25q    & \uline{0.618}  & 0.623                               & 0.674                          & 0.772           & \textbf{0.781}                                    \\
                                       
        & 75q    & 0.891                          & 0.894                               & \uline{0.883}  & 0.914           & \textbf{0.915}                           \\
\hline
      \multirow{5}{*}{\rotatebox[origin=c]{90}{\textbf{H95}}}                                 & Mean   & \uline{47.45}  & 43.95                               & \textbf{28.63}                 & 36.82           & 39.62                                    \\
                                       & s      & \uline{112.71} & 108.93                              & \textbf{90.68}                 & 106.06          & 109.85                                   \\
                                       & Median & \uline{2.83}   & \uline{2.83}        & 2.24                           & \textbf{2.00}      & \textbf{2.00}                               \\
                                       & 25q    & 1.41                           & 1.41                                &\uline{1.73}   & \textbf{1.00}      & \textbf{1.00}                               \\
         & 75q    & \uline{14.01}  & 11.53                               & 9.52                           & \textbf{3.30}             & 3.40                                      \\
\hline
      \multirow{5}{*}{\rotatebox[origin=c]{90}{\textbf{Sensitivity}}}                                 & Mean   & \uline{0.682}  & 0.690                                & 0.717                          & \textbf{0.756}           & 0.752                                    \\
                                       & s      & \uline{0.328}  & 0.324                               & \textbf{0.307}                 & 0.319           & 0.325                                    \\
                                       & Median & \uline{0.825}  & 0.827                               & 0.848                          & 0.886           & \textbf{0.890}                            \\
                                       & 25q    & \uline{0.595}  & 0.600                                 & 0.615                          & 0.756           & \textbf{0.758}                                    \\
 & 75q    & \uline{0.902}  & 0.905                               & 0.928                          & \textbf{0.951}           & 0.947                                    \\
\hline
    \multirow{5}{*}{\rotatebox[origin=c]{90}{\textbf{Specificity}}}                                   & Mean   & \uline{0.9997} & \uline{0.9997}      & \uline{0.9997} & \textbf{0.9998} & \textbf{0.9998}                          \\
                                       & s      & \textbf{0.0004}                & \textbf{0.0004}                     & \textbf{0.0004}                & \textbf{0.0004} & \textbf{0.0004}                          \\
                                       & Median & \uline{0.9998} & \uline{0.9998}      & \textbf{0.9999}                & \textbf{0.9999} & \textbf{0.9999}                          \\
                                       & 25q    & 0.9996                         & 0.9996                              & \uline{0.9995} & 0.9996          & 0.9996                                   \\
 & 75q    & \textbf{1.0000}                     & \textbf{1.0000}                          & \textbf{1.0000}                     & \textbf{1.0000}      & \textbf{1.0000}                              \\
\Xhline{2\arrayrulewidth}
\end{tabular}
\end{table}

\begin{figure}[ht!]
    \centering
    \includegraphics[width=0.75\columnwidth]{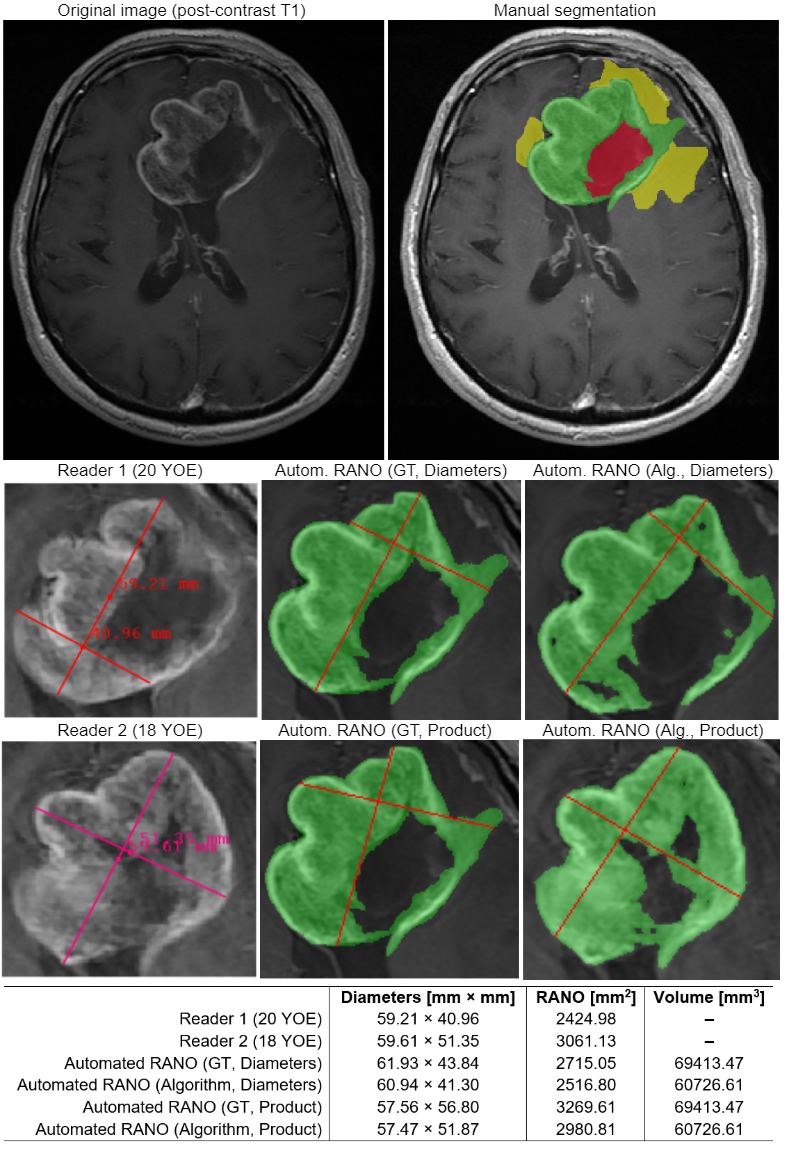}
    \caption{Qualitative and quantitative analysis can reveal important aspects concerning the behavior and abilities of the suggested pipeline---an example axial post-contrast T1 image and manual segmentation (green---ET, yellow---ED, red---surgical cavity). Below, the RANO bidimensional measurements for the two most experienced readers compared with Automated RANO (Diameters) and Automated RANO (Product) obtained from either the ground-truth or the algorithm’s segmentation. Manual calculation of RANO is subjective, difficult to reproduce and may easily lead to high inter-rater disagreement---the proposed algorithm offers full reproducibility and a more accurate optimization of the product diameters leading to a stronger correlation with the ET volume.}
    \label{fig:example1}
\end{figure}

The deep learning tumor segmentation was subsequently used to automatically obtain the tumor volumes and the bidimensional tumor measurements according to the Response Assessment in Neuro-Oncology (RANO) criteria. The results were then compared with measurements from experts. The tumor volumes extracted automatically were in almost perfect agreement with the values obtained by the readers (Figure~\ref{fig:bland_altman_et_ed_cavity}). For RANO, there was a significant disagreement across human raters which reflects the difficulty for radiologists to perform GBM tumor burden measurements, particularly in post-surgery settings, and the benefit that automation might bring. There were several indications that the deep learning model is more reliable and accurate than humans:
\begin{itemize}
    \item The algorithms had the highest agreements with the most expert radiologists (Figure~\ref{fig:bland_altman_auto_rano}).
    \item The pipeline returned higher bidimensional measurement than those reported by most human readers because the algorithm carefully finds the maximal enhancing tumor diameters (or their product) which cannot be done as reliably by a radiologist in a manual process (Figure~\ref{fig:example1}).
    \item The automated RANO correlated better with ground truth enhancing tumor volume than the RANO done by the readers (Figure~\ref{fig:rano_correlation}).
\end{itemize}
\noindent It is worth mentioning that optimizing the product of diameters instead of the maximum diameter was more effective at compensating for eventual errors in tumor segmentation and therefore provided more robust RANO measurement.

\begin{figure}[ht!]
    \centering
    \includegraphics[width=1\columnwidth]{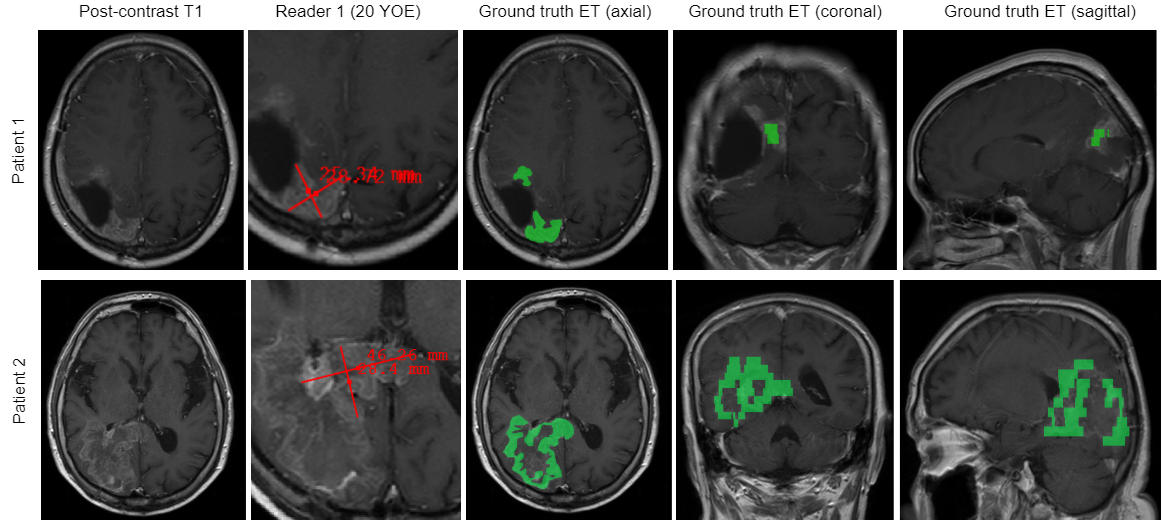}
    \caption{The examples of tumors with very different volume but with almost identical bidimensional RANO measurements, showing that in some cases, relying on 2D measurement can be misleading.}
    \label{fig:example2}
\end{figure}

The high correlation observed between RANO and tumor volume supports the concept of using bidimensional measurement for evaluating tumor burden in GBM patients. Nevertheless, while routinely applied, there are clearly limitations inherent to this approach as highlighted, for instance, by two patients in our dataset which have very similar RANOs (Patient A: 1071.45 mm$^2$, Patient B: 1137.73 mm$^2$) but very different tumor volumes (Patient A: 8502.33 mm$^3$, Patient B: 66287.65 mm$^3$). As this example shows, analyzing a 2D MRI slice is not always sufficient to accurately quantify tumor burden (Figure~\ref{fig:example2}). Volumetric tumor segmentation, which requires automation for its integration into the clinical workflow, is also a necessary initial step toward more advanced image analysis that can be used to identify novel biomarkers---this includes intensity-related features, 2D and 3D shape characteristics, texture analysis of the tumor tissue~\citep{10.1007/978-3-319-11331-9_53}, as well as other radiomic parameters~\citep{AYADI2019144} that can be used to build predictive models in diagnosis, prognosis and therapeutic response~\citep{10.3389/fonc.2019.00374,Park2020,Suter2020}.

\section{Conclusions and Future Work}\label{sec:contributions}

For patients with glioblastoma, the evaluation of tumor burden and response to therapy is based on the bidimensional measurement of the T1 contrast enhancing tumor area and the qualitative evaluation of abnormalities on T2/FLAIR MRI scans. As gliomas are often very heterogeneous in appearance and shape, this assessment is complex to perform and associated with high variability, particularly in the post-surgery setting. In this work, we approached this issue and introduced an easy to use, built upon and expanding the recent advances in the machine learning field, and thoroughly validated pipeline that automatically segments pre- and post-operative MRIs from GBM patients and delineates tumor sub-regions, including contrast enhancing area, T2/FLAIR hyperintensities and surgical cavity. The entire process, with volumetric and RANO calculations, is executed in a fraction of time needed by humans with performance matching or exceeding radiologists. The algorithm’s potential for improving and simplifying radiological assessment of glioma patients opens the door to its deployment in clinical trials---as shown in our experimental study performed over a very large number of patients---and its integration into the clinical workflow. The automatic measurements were reliable, fast and in statistically significant agreement with the most experienced radiologists. Finally, we introduced a rigorous manual delineation process that was followed by radiologists to provide ground-truth segmentations, together with additional information related to their confidence and analysis time.

The results reported in this paper constitute an interesting departure point for further research. The follow-up steps for our segmentation algorithm would be to evaluate its robustness for longitudinal assessment and its ability to adequately identify disease progression or response during the course of a number of drug treatments. Additionally, it would be interesting to compare its performance with the top-performing techniques from BraTS 2021 once they are published. Having a comprehensive and objective comparative study of the best segmentation techniques in the literature would be an important step toward building robust and certified AI-powered tools that could be utilized in clinical practice. Since gathering more clinical MRI data with the corresponding ground-truth data is extremely costly, training the algorithms in the federated learning approach bloomed as an exciting opportunity as well, as it could help us train the deep learning models over massively large training samples without the necessity of sharing them across institutions~\citep{Sheller2020}. Finally, the improvement of our deep learning pipeline might also be possible by evaluating the image features associated with poorer  performance (e.g.,~for low-quality or deformed MRI scans), and building data augmentation routines that focus more specifically on these characteristics~\citep{10.3389/fncom.2019.00083}. 

\section*{Acknowledgements}
The authors thank Josep Garcia and Yannick Kerloegen (Hoffmann-La-Roche) for helping us using the GBM MRI data from the clinical treatment trial as well as Marek Pitura and Daria Bernys (Future Processing Healthcare) for their valuable help in managing this study.

\bibliography{mybibfile}

\end{document}